\documentclass[11pt,runningheads]{llncs}
\usepackage[T1]{fontenc}
\usepackage{booktabs,tabularx,makecell}
\usepackage{newtxtext}
\usepackage[varvw]{newtxmath}

\usepackage{graphicx}
\usepackage{cite}

\usepackage[
    left=1in,
    right=1in,
    top=1in,
    bottom=1in
]{geometry}

\usepackage{algorithm}
\usepackage{algpseudocode}

\newtheorem{observation}[lemma]{Observation}

\usepackage[colorlinks=true, linkcolor=blue!60!black, citecolor=blue!60!black, urlcolor=blue!60!black]{hyperref}
\usepackage[nameinlink]{cleveref}

\usepackage{thmtools}
\crefname{lemma}{Lemma}{Lemmas}
\crefname{theorem}{Theorem}{Theorems}
\crefname{section}{Section}{Sections}
\crefname{corollary}{Corollary}{Corollaries}
\crefname{figure}{Figure}{Figures}
\crefname{remark}{Remark}{Remarks}
\crefname{definition}{Definition}{Definitions}
\crefname{algorithm}{Algorithm}{Algorithms}
\crefname{instance}{Instance}{Instances}
\crefname{observation}{Observation}{Observations}
\crefname{appendix}{Appendix}{Appendices}
\crefname{subappendix}{Appendix}{Appendices}
\crefname{subsubappendix}{Appendix}{Appendices}

\usepackage{xspace}

\newcommand{\alg}{\textsc{Alg}\xspace}
\newcommand{\opt}{\textsc{Opt}\xspace}

\definecolor{job1}{HTML}{56b3e9}
\definecolor{job2}{HTML}{e69f00}
\definecolor{job3}{HTML}{009e74}
\definecolor{job4}{HTML}{cc79a7}
\definecolor{job5}{HTML}{d55e00}
\definecolor{job6}{HTML}{0071b2}

\RequirePackage{tikz}
\usetikzlibrary{calc,shapes,shapes.symbols,shapes.geometric,positioning,fit,arrows,patterns,decorations.pathreplacing}
\tikzset{every node/.style={anchor=center}}

\begin{document}

\title{Speed-Robust Scheduling for Total Completion Time}

\author{Moritz Buchem \and Nicole Megow \and Morten Weber}

\authorrunning{Buchem, Megow \& Weber}

\institute{Faculty of Mathematics and Computer Science, University of Bremen, Germany\\
\email{\{mbuchem,nicole.megow,moweber\}@uni-bremen.de}}

\maketitle

\begin{abstract}

We study two-stage speed-robust scheduling for minimizing total completion time. Jobs must be partitioned into $M$ bags before an adversary reveals the speeds of at most $M$ machines. 
Each bag is then assigned to one available machine, and the jobs on each machine may be processed in any order independent of the bags. We compare with an optimal schedule that knows the speeds in advance and assigns jobs individually. This model has been studied for makespan minimization. We initiate its study for total completion time, as suggested by Stein and Zhong (TALG 2020).

For an unknown number of identical machines, we show that the optimal deterministic robustness factor is exactly~$13/12$, both for unit-size jobs and for arbitrary processing times. For infinitesimally small jobs, the factor lies between $31/30\approx1.0333$ and $28/27\approx1.0370$, well below the optimal makespan factor of about $1.2071$.
General speeds make the problem strictly harder: no deterministic algorithm is better than $1.2669$-robust, already for infinitesimally small jobs. We give a $25/16$-robust algorithm for arbitrary processing times and improve the guarantee to $1681/1152\approx1.4592$ for infinitesimally small jobs. Our main tools are a layer decomposition that extends guarantees from unit-size jobs to arbitrary processing times when one assignment satisfies them for every prefix in non-increasing processing-time order, and a threshold analysis of the bag counts produced by greedy assignment rules.
\end{abstract}

\section{Introduction}
\label{sec:intro}
We study two-stage scheduling for minimizing the total completion time when the available machines and their speeds are initially unknown. The jobs and their processing times are known in advance, but the jobs must be grouped before the machine environment is revealed. Such grouping may be necessary when computational tasks and their input data are bundled before submission to a shared computing system. The challenge is to choose groups that admit good schedules for every possible machine environment.

We consider \emph{speed-robust scheduling}, which was introduced by Stein and Zhong~\cite{DBLP:journals/talg/SteinZ20} for an unknown number of identical machines and extended to general speeds by Eberle et al.~\cite{DBLP:journals/mp/EberleHMNSS23}. In Phase~I, an algorithm partitions the jobs into~$M$ groups, called \emph{bags}. An adversary then reveals either a number $1\le m\le M$ of identical machines (\emph{speeds in~$\{0,1\}$}) or a nonnegative speed for each of the $M$ machines (\emph{general speeds}). Machines with speed zero are unavailable. In Phase~II, all jobs of a bag must be assigned to the same available machine, and several bags may share a machine. Jobs from different bags may be interleaved and are processed in shortest-processing-time-first (SPT) order, which minimizes the total completion time for a fixed assignment~\cite{Smith56}.

We evaluate an algorithm by its \emph{robustness factor}, the worst-case ratio between the objective value of its schedule and that of an optimal schedule that knows the machine environment in advance and may assign jobs individually. This comparison measures the loss caused by fixing the bags before the machine specification is known. Following Eberle et al.~\cite{DBLP:journals/mp/EberleHMNSS23}, we distinguish infinitesimally small jobs (\emph{sand}), unit-size jobs (\emph{bricks}), and jobs with arbitrary processing times (\emph{rocks}).

Previous work on this model has focused primarily on makespan minimization. For an unknown number of identical machines, Stein and Zhong~\cite{DBLP:journals/talg/SteinZ20} gave the first algorithms and lower bounds, including a $(5/3+\varepsilon)$-robust algorithm for rocks. Eberle et al.~\cite{DBLP:journals/mp/EberleHMNSS23} closed the gaps for sand and bricks, establishing optimal uniform robustness factors of~$(1+\sqrt2)/2\approx1.2071$ and $4/3$, respectively. For general speeds, they determined the optimal factor for sand, which tends to $e/(e-1)\approx1.5820$ as the number of machines grows, and gave constant guarantees for bricks and rocks; the guarantee for bricks was subsequently improved by Mina\v{r}\'{\i}k and Sgall~\cite{DBLP:conf/approx/MinarikS24}. For identical machines, Yu and Tan~\cite{DBLP:conf/cocoa/YuT24} gave a simpler optimal algorithm for bricks.

Stein and Zhong~\cite{DBLP:journals/talg/SteinZ20} explicitly raised total completion time as an open problem, noting that it introduces additional questions about job ordering. Unlike makespan, this objective depends on the processing order on each machine. Although SPT determines the optimal processing order for a fixed assignment, the bags must still admit an assignment of small total completion time for every revealed machine environment.
For sand, the objective reduces to a sum of squared assigned volumes weighted by inverse speeds. Every $\rho$-robust makespan algorithm for sand is also $\rho$-robust for total completion time~(\Cref{lemma:Makespan_To_MinSum}). Thus, the makespan factors $1.2071$ and $1.5820$ provide natural benchmarks.
Our results show that substantially smaller robustness factors are possible.

\begin{table}[t]
\centering
\setlength{\tabcolsep}{3pt}
\newcommand{\boundref}[2]{
  \makecell[c]{$#1$\\{\small\Cref{#2}}}}
\begin{tabular*}{\textwidth}{@{\extracolsep{\fill}}lcccc@{}}
\toprule
& \multicolumn{2}{c}{Speeds in $\{0,1\}$}
& \multicolumn{2}{c}{General speeds}\\
\cmidrule(lr){2-3}\cmidrule(l){4-5}
Jobs & Lower bound & Upper bound & Lower bound & Upper bound\\
\midrule
Sand
  & \boundref{31/30\approx1.0333}{theorem:Sand_LB_1_033}
  & \boundref{28/27\approx1.0370}{theorem:Sand_UB_Improved}
  & \boundref{1.2669}{thm:speeds-sand-lower}
  & \boundref{1681/1152\approx1.4592}{thm:speeds-imbalanced-sand}\\
\addlinespace[5pt]
Bricks and rocks
  & \multicolumn{2}{c}{
      \boundref{13/12\approx1.0833}{thm:LB_Bricks,thm:UB_Bricks_Rocks_13/12}}
  & \boundref{1.2669}{thm:speeds-sand-lower}
  & \boundref{25/16=1.5625}{thm:speeds-cyclicbags-2516}\\
\bottomrule
\end{tabular*}
\setlength{\abovecaptionskip}{5pt}
\caption{Summary of our robustness bounds (uniform over the number of bags). Entries spanning two columns are tight.} 
\label{tab:Results}
\end{table}

\subsection{Our results}\label{sec:results}
We initiate the study of speed-robust scheduling for total completion time and obtain tight or nearly tight bounds for an unknown number of identical machines and small constant bounds for general speeds. Our bounds are uniform over the number~$M$ of bags: upper bounds hold for every~$M$, while lower bounds rule out smaller guarantees valid for every~$M$. All upper bounds are attained by deterministic polynomial-time algorithms and lower bounds assume deterministic algorithms. \Cref{tab:Results} summarizes the bounds. 

\smallskip\noindent\textbf{Speeds in $\{0,1\}$\ } The optimal robustness factor is exactly $13/12$ for both bricks and rocks (\Cref{thm:LB_Bricks,thm:UB_Bricks_Rocks_13/12}). Thus, arbitrary processing times are no harder than unit-size jobs. For the makespan, in contrast, rocks are not known to be as easy as bricks. Our algorithm deliberately creates bags with different job counts, which can be combined to balance the machines in Phase~II. For bricks, the analysis controls the variance of the machine job counts. For rocks, we construct the bags in non-increasing order of processing times. A single bag-to-machine assignment satisfies the brick guarantee for every prefix of this order viewed as unit-size jobs.
A layer decomposition then transfers the guarantee to the original processing times~(\Cref{lem:prefix-transfer}).

For sand, the optimal robustness factor lies between \(31/30\) and \(28/27\), leaving a gap of \(1/270<0.004\). The lower bound holds even for randomized algorithms against an oblivious adversary.
Equal-volume bags achieve a tight factor of $9/8$ (\Cref{theorem:Sand_Equally_UB}); our bag volumes instead follow a linear profile, so that small bags can be paired when fewer machines are available.
Our lower bound combines Cauchy--Schwarz relaxations for four machine counts with  $M=12$ bags.

\smallskip\noindent\textbf{General speeds\ } Allowing arbitrary speeds makes the problem strictly harder: no deterministic algorithm is better than~$1.2669$-robust, already for sand (\Cref{thm:speeds-sand-lower}). This exceeds our factor of $13/12$ even for rocks for~$\{0,1\}$-speeds, and by \Cref{obs:sand-to-bricks} the bound carries over to bricks and rocks. The adversary uses one fast machine and $M-1$ equally slow machines, choosing their speeds after observing the bags. A fractional cost profile captures the trade-off between sending bags to the slow machines and concentrating volume on the fast machine. The restriction to $M$ bags bounds the integral of its reciprocal slope, independently of the bag volumes. This limits how well any partition can accommodate all speed vectors.

For the upper bound, we give a common $25/16$-robust algorithm for sand, bricks, and rocks (\Cref{thm:speeds-cyclicbags-2516}). It uses equal-volume bags for sand and assigns jobs cyclically to bags in non-increasing processing-time order for bricks and rocks. In Phase~II, bags are assigned by a greedy rule based on bag counts and machine speeds.
Its threshold structure keeps bag counts proportional to machine speeds up to rounding. For bricks, the analysis compares machine costs with those of an optimal schedule using correction terms whose sum is non-positive. The prefix transfer again handles rocks.
The factor~$25/16$ is tight for these bag constructions, even with an optimal Phase~II assignment (\Cref{thm:speeds-cyclicbags-2516}). For sand, bags of different volumes and a modified Phase~II rule improve the guarantee to $1681/1152\approx1.4592$ (\Cref{thm:speeds-imbalanced-sand}), clearly below the makespan benchmark of~$1.5820$. Smaller bags allow slow machines to receive less volume. We assign them to machines with less speed per assigned bag. Cauchy--Schwarz bounds the cost by a sum of squared bag volumes, each weighted by the number of bags per unit of speed on its machine. We use threshold inequalities of the assignment rule and convexity to bound this sum.

\smallskip\noindent\textbf{Additional variants\ }
We show that the freedom to reorder jobs is essential: in a non-adaptive variant, where every bag must be processed contiguously, no algorithm with~$M$ bags is better than $M$-robust, already for identical machines. This bound is trivially achieved by placing all jobs in a single bag; we determine the optimal robustness factor exactly as a function of $M$ and the number of distinct processing times (\Cref{app:non-adap-detail}).

We also study a model proposed by Liu, Zhang, and Zhao~\cite{DBLP:conf/ijcai/LiuZZ26}, in which jobs are assigned directly to machines in Phase~I and the adversary chooses which machines fail. Jobs on surviving machines must remain there. For identical machines, we give an algorithm matching the unrestricted optimum when jobs from failed machines may be reassigned individually.
We also introduce a more restrictive variant in which they must be reassigned together, and give bounds for it (\Cref{app:remarks-proofs}).

\subsection{Further related work}
Without bag constraints, the total completion time can be minimized in polynomial time if the machine environment is known: by SPT list scheduling on identical machines~\cite{DBLP:journals/cacm/BrunoCS74} and by a greedy assignment to positional slots on uniformly related machines~\cite{10.1287/opre.21.3.846}. For a single machine whose speed changes in an unknown way, Epstein et al.~\cite{DBLP:journals/siamcomp/EpsteinLMMMSS12} constructed universal job sequences for the total weighted completion time that are good for all speed patterns simultaneously.

Other work assumes additional information about the machine environment.
Buchem et al.~\cite{DBLP:conf/approx/BuchemERSW24} considered a stochastic number of identical machines with a known distribution and gave PTASs for the expected makespan and the expected minimum machine load.
Epstein and Levin~\cite{DBLP:conf/stacs/EpsteinL25} improved these to EPTASs and also considered expected~$\ell_p$ norms of the loads.
Their guarantees are relative to the best two-stage solution for the given distribution, whereas our benchmark may assign jobs individually. Balkanski et al.
\cite{DBLP:journals/mst/BalkanskiOSW25} considered predicted machine speeds and obtained makespan guarantees that improve with prediction accuracy while remaining bounded for arbitrary predictions.

Epstein~\cite{DBLP:journals/disopt/Epstein26} studied an online variant of speed-robust scheduling in which jobs arrive one by one and must be assigned to bags immediately, for makespan minimization and for maximizing the minimum load. She also studied in~\cite{DBLP:journals/cejor/Epstein26} ordinal scheduling with an unknown number of machines, where processing times remain unknown when bags are formed and merged.

\section{Preliminaries}\label{sec:prelim}
This section formalizes the model from the introduction and introduces three auxiliary results used throughout the paper: a transfer of guarantees from bricks to rocks, a transfer of lower bounds from sand to bricks and rocks, and a transfer of makespan guarantees to the total completion time for sand. Omitted proofs are in~\Cref{app:prelim-proofs}.

\medskip\noindent\textbf{Problem definition\ }
An instance consists of $n$ jobs with processing times $p_j>0$, $j\in[n]$, and the number~$M$ of bags, which is also the maximum number of machines. All jobs are available at time zero. In Phase~I, an algorithm partitions the given jobs into bags $B_1,\ldots,B_M$, some of which may be empty. Then, an adversary reveals a speed vector $s\in\mathbb{R}_{\ge0}^M\setminus\{0\}$: a job~$j$ on a machine of speed $s_i>0$ takes $p_j/s_i$ time units, and a machine of speed zero is unavailable. Let $m:=|\{i\in[M]:s_i>0\}|$, and relabel the machines so that the available machines are indexed by~$1,\ldots,m$. We call the unrestricted choice of $s$ the setting of \emph{general speeds}. For \emph{speeds in $\{0,1\}$}, the adversary chooses only the number $m\in[M]$ of available identical machines.

In Phase~II, the algorithm assigns every bag to one of the available machines via a map $\pi\colon[M]\to [m]$ where several bags may share a machine. Machine $i$ then processes the jobs $J_i:=\bigcup_{r\colon\pi(r)=i}B_r$ in non-decreasing order of processing times (SPT), one at a time without preemption or idle time.

\medskip\noindent\textbf{Objective and robustness guarantee\ }
The objective is to minimize the total completion time $\sum_{j\in[n]}C_j$, where $C_j$ is the completion time of job~$j$. For an instance~$I$ and a speed vector~$s$, let $\alg(I,s)$ denote the cost of the algorithm's schedule, and let $\opt(I,s)$ denote the minimum total completion time of a schedule that knows $s$ in advance and may assign jobs individually, without bag constraints. An algorithm is \emph{$\rho$-robust} if $\alg(I,s)\le\rho\,\opt(I,s)$ for every instance~$I$ and every feasible speed vector~$s$. 
Its \emph{robustness factor} is the infimum over all such~$\rho$. The \emph{optimal robustness factor} of a setting is the infimum of the robustness factors of all deterministic algorithms. For a fixed number $M$ of bags, we define the corresponding robustness factors by restricting to instances with $M$ bags. We omit the instance $I$ from the notation whenever it is clear.

\medskip\noindent\textbf{Sand, bricks, and rocks\ } 
For sand, we normalize the total volume to one. Let $b_r$ denote the volume of bag~$B_r$ and~$y_i:=\sum_{r:\pi(r)=i}b_r$ the volume assigned to machine~$i$. We view sand as the limit of equal-size jobs whose common size~$\delta$ tends to zero. For fixed assigned volumes, total completion time grows like $1/\delta$. We therefore multiply algorithmic and optimal costs by $\delta$, preserving their ratio and obtaining a finite limit. If machine~$i$ receives $k_i$ jobs of total volume~$y_i=k_i\delta$, its scaled total completion time is
\[
 \delta\sum_{h=1}^{k_i}\frac{h\delta}{s_i}
 =\frac{y_i^2+\delta y_i}{2s_i}
 \longrightarrow\frac{y_i^2}{2s_i}
 \qquad\text{as }\delta\to0.
\]
We therefore use $\frac12\sum_i y_i^2/s_i$ as the objective for sand. Write $S:=\sum_{i=1}^m s_i$. By Cauchy--Schwarz, $\sum_{i=1}^m y_i^2/s_i\ge(\sum_{i=1}^m y_i)^2/S=1/S$, with equality for $y_i=s_i/S$. Hence, the unrestricted optimum for sand has cost $1/(2S)$, and the ratio between the cost of an assignment with loads~$y_i$ and this optimum is
\[
  R(s):=S\sum_{i=1}^m\frac{y_i^2}{s_i}.
\]

For speeds in $\{0,1\}$, we write $R(m):=m\sum_{i=1}^m y_i^2$ for this ratio.
For bricks, a machine of speed~$s_i$ processing $k$ unit-size jobs contributes $F(k)/s_i$, where $F(k):=k(k+1)/2$.
For speeds in $\{0,1\}$, the linear part of the total cost is always $n/2$; for general speeds, it depends on the assignment.

\medskip\noindent\textbf{Transferring guarantees\ }
Write $p_1\ge\cdots\ge p_n$ and set $p_{n+1}=0$. For any fixed assignment, its cost under SPT is~a weighted sum of the costs of treating each prefix as bricks, with weight $p_r-p_{r+1}$ for the prefix~$1,\ldots,r$. Thus, a guarantee for bricks transfers to rocks if the same assignment satisfies it for every prefix.

\begin{restatable}{lemma}{restatePrefixTransfer}\label{lem:prefix-transfer}
Fix an instance~$I$ with $p_1\ge\cdots\ge p_n$ and speeds~$s$. Consider an algorithm's assignment $A$ on~$(I,s)$. Suppose that, for every $r\in[n]$, restricting $A$ to the jobs $1,\ldots,r$ and setting their processing times to one gives~a schedule of cost at most $\alpha$ times the unrestricted optimum for $r$ unit-size jobs and speeds~$s$. Then $\alg(I,s)\le\alpha\,\opt(I,s)$.
\end{restatable}

Lower bounds for sand also extend to bricks and rocks by a limiting argument: sufficiently many unit-size jobs approximate sand after normalizing the total volume. 

\begin{restatable}{observation}{restateSandToBricks}\label[observation]{obs:sand-to-bricks}
For every fixed number of bags, every lower bound on the optimal robustness factor for sand also holds for bricks and rocks in the same machine setting.
\end{restatable}

For sand, robustness guarantees for the makespan carry over to our objective.

\begin{restatable}{lemma}{restateMakespanToMinSum}
\label{lemma:Makespan_To_MinSum}
For sand, every algorithm that is $\gamma$-robust for the makespan is also $\gamma$-robust for the total completion time, using the same bags and Phase~II assignment.
\end{restatable}

\section{Speeds in $\{0,1\}$}\label{sec:Equal_Speeds}
We first consider speeds in $\{0,1\}$, where the adversary only chooses the number $m\leq M$ of available machines, all running at unit speed.
Since SPT is optimal for $m=1$, and assigning jobs cyclically in non-increasing order of size to the bags is optimal for $M=2$, the challenge arises for $M\geq3$: the bags should be nearly balanced for $m=M$, yet still allow balanced assignments when fewer machines are available.
For sand, we resolve this trade-off up to an additive~$0.004$: bag volumes following a linear profile yield a robustness factor of $28/27$, and no algorithm achieves a factor smaller than~$31/30$~(\Cref{theorem:Sand_UB_Improved,theorem:Sand_LB_1_033}).

For bricks and rocks, we give a $13/12$-robust algorithm and show that no deterministic algorithm does better (\Cref{thm:UB_Bricks_Rocks_13/12,thm:LB_Bricks}).
Omitted proofs are provided in~\Cref{app:equal-speeds-proofs}.

\subsection{Sand: linear bag profiles within $0.004$ of optimal}
For sand, we normalize the total volume to one, so that $\sum_{i=1}^M b_i=1$, and denote the resulting machine loads by~$y_1,\ldots,y_m$. By \Cref{sec:prelim}, the ratio of a bag assignment on~$m$ machines is $R(m):=m\sum_{i=1}^my_i^2$.
A simple algorithm already beats the makespan benchmark of $1.2071$ (cf.~\Cref{lemma:Makespan_To_MinSum}): If all bags receive volume $1/M$ and Phase~II distributes the bags as evenly as possible, then $R(m)=1+r(m-r)/M^2\leq 9/8$, where~$M=qm+r$ with $0\leq r<m$. This is tight for $M=4$ and $m=3$ (\Cref{theorem:Sand_Equally_UB} in App.~\ref{app:sand-equal}).
For~$M/2<m<M$, equal bags lose because some machines must receive two bags and thus carry twice the load of others. Unequal bag volumes reduce this imbalance: small bags can be paired when fewer machines are available, while large bags remain~alone.

Our algorithm uses bag volumes that follow a linear profile, ranging approximately from $2/(3M)$ to~$4/(3M)$.

\begin{restatable}{theorem}{restateSandLinear}\label{theorem:Sand_UB_Improved}
For sand and speeds in $\{0,1\}$, the linear-profile algorithm has an asymptotically tight robustness factor of $28/27\approx1.0370$.
\end{restatable}
\begin{proof}[sketch]
In Phase~I, we use the linear profile $x(t)=2(1+t)/3$, defining the bag volumes by $b_i=\int_{(i-1)/M}^{i/M}x(t)\,dt$.
For $m>M/2$, we pair the $2(M-m)$ smallest bags in opposite order, i.e., bag $i$ with bag $2(M-m)+1-i$, and assign the remaining bags individually.
Since the bag volumes form an arithmetic progression, all pairs have equal total volume. Then, Jensen's inequality bounds the contribution of the remaining bags, yielding an upper bound on $R(m)$ that depends only on the fraction of bags that are paired. We show that this bound is at most $28/27$.

For $m\leq M/2$, choose the smallest power of two $h$ with $hm>M/2$. We first assign the bags to $hm$ virtual machines as above and then merge groups of $h$ machines. By Cauchy--Schwarz, the resulting ratio does not exceed that of the virtual assignment.\qed
\end{proof}

For the lower bound, we use a fixed probability distribution over four machine counts with $M=12$ bags. Cauchy--Schwarz yields an expected ratio of at least $31/30$ for all bag constructions. By Yao's principle the bound also holds for randomized algorithms.
\begin{restatable}{theorem}{restateSandLB}\label{theorem:Sand_LB_1_033}
    For sand and speeds in $\{0,1\}$, no algorithm for minimizing the total completion time achieves a robustness factor smaller than $31/30\approx1.0333$.
\end{restatable}

\subsection{Bricks and rocks: a tight bound of $13/12$}
We construct an algorithm for bricks whose guarantee holds simultaneously for every prefix of the sorted job sequence. \Cref{lem:prefix-transfer} allows us to extend the guarantee to rocks.

Our algorithm \textit{$3{:}2$-Cyclic}
sorts the jobs in non-increasing order of processing times.
For $M\leq2$, it assigns jobs cyclically to the bags and the bags optimally to the available machines. Scheduling the jobs in SPT order is then optimal, as discussed above.
For $M\geq3$, let $L:=\lfloor M/2\rfloor$ and $x:=2M+L=\lfloor5M/2\rfloor$.
In Phase~I, we partition this sequence into consecutive packages of $x$ jobs, with possibly fewer jobs in the last package. Within each package, we assign the jobs cyclically to the bags, starting with~$B_1$.
Thus, each complete package contributes three jobs to each of the first $L$ bags and two jobs to each of the remaining $M-L$ bags. We call these \textit{long} and \textit{short bags}, respectively.

In Phase~II, let $K:=m\lceil L/m\rceil$. We assign $B_1,\ldots,B_K$ cyclically to the machines, starting with machine~$1$. Thus, all long bags are assigned, and every machine receives exactly $\lceil L/m\rceil$ bags. The remaining bags are assigned in increasing bag-index order, cyclically in reverse machine order, starting with machine~$m$. All bag labels are assigned, including those of empty bags. On each machine, we schedule the jobs in SPT order. A formal description and an illustration of both phases are given in \Cref{app:Bricks_App}.

The $3{:}2$ pattern is the discrete version of a two-size construction for sand: volumes $3/(2M+L)$ for the first $L$ bags and $2/(2M+L)$ for the others, assigned by longest-processing-time-first (LPT) in Phase~II, give a tight robustness factor of $18/17$ for sand. For bricks, the $3{:}2$-Cyclic algorithm is asymptotically $18/17$-robust as $n\to\infty$ for every fixed~$M$ (\Cref{theorem:Sand_2_Sizes,rem:bricks-asymptotic} in \Cref{app:sand-two-sizes}).
We use this pattern rather than the linear profile because it is easy to realize with integral bag loads and is consistent on prefixes of the job sequence, which we need for rocks. Since the lower bound of $13/12$ for bricks exceeds both $18/17$ and~$28/27$, a finer profile cannot improve the worst case.

For bricks, let $b_i:=|B_i|$ denote the load of bag~$B_i$, and let $y_i$ denote the number of jobs assigned to machine~$i$. We write \(\alg(m,M)\) and \(\opt_m\) for the algorithm’s cost and the unrestricted optimum on \(m\) identical machines, respectively.

\begin{restatable}{theorem}{restateBricksRocksUB}\label{thm:UB_Bricks_Rocks_13/12}
    For speeds in $\{0,1\}$, the \textit{$3{:}2$-Cyclic} algorithm is $13/12$-robust for minimizing the total completion time of bricks and rocks.
\end{restatable}

\begin{proof}[sketch]
We first consider bricks with $M\geq3$. The forward cyclic rounds yield non-increasing machine loads, while the reverse rounds add non-decreasing loads. Consequently, the machine loads differ by at most $\max\{1,s\}$, where~$s:=b_{L+1}$ is the largest short-bag load.

For $s\leq2$, the schedule is either optimal or has loads in $\{\beta,\beta+1,\beta+2\}$ with $\beta\geq2$. In the latter case, balancing each pair of extreme loads saves one unit of cost, while the balanced pair contributes at least $12$, giving the factor~$13/12$. The same argument handles any later case with these three consecutive loads.

For the remaining cases, let $y_i$ denote the load of machine~$i$, and write $\overline y:=n/m$ and $V:=m^{-1}\sum_{i=1}^m(y_i-\overline y)^2$. Since~$F(k)=k(k+1)/2$, the algorithm's cost is $m(V+\overline y^2+\overline y)/2$. Convexity bounds the optimum from below by~$m(\overline y^2+\overline y)/2$, so
\[
 \frac{\alg(m,M)}{\opt_m}
 \leq1+\frac{V}{\overline y^2+\overline y}.
\]
It therefore suffices to show $12V\leq\overline y^2+\overline y$. For this, we distinguish whether a short bag is assigned alone. If not, every machine receives a long bag or at least two short bags. Thus, the minimum machine load $\beta$ satisfies $\beta\geq3s/2-1$, and all machine loads belong to $[\beta,\beta+s]$. If a short bag is assigned alone, all long bags are assigned alone as well. Let $a$ be the smallest load of a short bag assigned alone. All machine loads belong to~$[a,2a]$, and the loads of the machines receiving long bags differ from $3a/2$ by at most one. These machines account for at least one third of all available machines.

In both cases, we bound the squared deviations from the midpoint $c$ of the respective interval and use
\[
 V=\frac1m\sum_{i=1}^m(y_i-c)^2-(\overline y-c)^2.
\]
The resulting bounds yield $12V\leq\overline y^2+\overline y$ for the remaining cases. Details are given in \Cref{app:Bricks_App}.
The package boundaries and the assignment of bag labels are independent of $n$ and the processing times. Thus, every prefix of the sorted job sequence is assigned as in the corresponding bricks instance, and \Cref{lem:prefix-transfer} extends the guarantee to rocks.
\qed
\end{proof}

We next show that the robustness factor of $13/12$ is best possible, even for bricks.
\begin{restatable}{theorem}{restateBricksLB}\label{thm:LB_Bricks}
    For speeds in $\{0,1\}$, no deterministic algorithm for minimizing the total completion time of bricks achieves~a robustness factor smaller than $13/12$.
    For rocks, the same lower bound holds for every~$n\geq6$. 
\end{restatable}

\begin{proof}
    Consider $n=6$ unit jobs and $M=3$ bags. If the bag loads are balanced, each bag contains two jobs. The adversary chooses $m=2$, for which the best possible machine loads for the algorithm are~$(4,2)$.
    Thus,~$\alg(2,3)/\opt_2\geq13/12$.
    Otherwise, the adversary chooses~$m=3$.
    Since  some bag contains at least three jobs, the algorithm's cost is at least that of machine loads $(3,2,1)$.
    Hence, $\alg(3,3)/\opt_3\geq10/9>13/12$.

    For rocks and every $n\geq7$, adding $n-6$ jobs of size $\varepsilon>0$ and letting $\varepsilon\to0$ yields the same lower bound (\Cref{app:rocks-lb}).
    \qed
\end{proof}

\section{General Speeds}
\label{sec:speeds}
We now study general speeds, where the adversary chooses the individual machine speeds rather than just the number of available machines. We show that general speeds are strictly harder (\Cref{thm:speeds-sand-lower}), give a common $25/16$-robust algorithm for sand, bricks, and rocks that is tight for its bag constructions (\Cref{thm:speeds-cyclicbags-2516}), and improve the guarantee for sand using bags of different volumes (\Cref{thm:speeds-imbalanced-sand}).

Throughout this section, we normalize the total speed to one, which does not affect robustness factors. Ties are broken according to a fixed order on the machines. Omitted proofs can be found in \Cref{app:speeds-proofs}.

\subsection{General speeds are strictly harder}
\label{sec:speed-sand}
We first present a lower bound for sand that holds even with an optimal Phase~II assignment. For $M\ge 2$, given the bag volumes $b_1,\ldots,b_M$ chosen in Phase~I, the adversary chooses $M-1$ slow machines of speed~$\sigma/(M-1)$ and one fast machine of speed $1-\sigma$, where $0<\sigma\le(M-1)/M$. Thus, $\sigma$ is the total speed of the slow machines.

Assigning bags to slow machines reduces the volume processed by the fast machine, but incurs a cost depending on the squared volumes of these bags. The adversary chooses the total speed $\sigma$ of the slow machines to exploit this trade-off. 

\begin{restatable}{theorem}{restateSpeedsSandLB}
\label{thm:sand-lower}\label{thm:speeds-sand-lower}
For general speeds, no deterministic algorithm for minimizing the total completion time of sand achieves~a robustness factor smaller than $\gamma=(1-\frac12e^{-3/2})^{-2}>1.2669$, uniformly over the number of machines.
\end{restatable}

\begin{proof}[sketch]
 Fix the bag volumes chosen in Phase~I. For a prescribed total volume~$u$ assigned to the slow machines, we relax the choice of bags by allowing fractional selections. Let $q(u)$ be the minimum of $(M-1)\sum_r z_rb_r^2$ subject to $\sum_r z_rb_r=u$ and~$0\le z_r\le1$. 
 Since selecting a bag of positive volume $b_r$ has a relaxed cost of~$(M-1)b_r$ per unit of selected volume, the minimum is attained by selecting bags in non-decreasing order of volume with at most one bag selected fractionally. Dropping cross terms between bags assigned to the same slow machine and applying this relaxation shows that every Phase~II assignment that sends a total volume of $u$ to the slow machines has ratio at least
 \[
  \frac{q(u)}{\sigma}+\frac{(1-u)^2}{1-\sigma}.
 \]
 The function $q$ is convex and each positive bag volume $b_r$ contributes a segment of length $b_r$ and slope $(M-1)b_r$. Hence, 
 \[
  \int_0^1\frac{du}{q'(u)}
  =\frac{|\{r:b_r>0\}|}{M-1}
  \le\frac{M}{M-1}.
 \]
 Thus, each positive bag contributes exactly $1/(M-1)$, independently of its volume.
 
 Suppose that the fixed bags admit a ratio of at most $\rho$ for every speed vector. We show that this requires~$q(u)\le(u+\sqrt{\rho}-1)^2$ for every $u$. If this inequality fails, we construct a speed vector for which every assignment has ratio above $\rho$, using either uniform speeds or a tangent to $q$. Using convexity and $q(0) = 0$, we show that this upper bound on $q$ implies that the reciprocal-slope integral is larger than $M/(M-1)$ whenever $\rho < \gamma$ and $M$ is sufficiently large. This contradicts the upper bound on the integral above. \qed
 \end{proof}
By \Cref{obs:sand-to-bricks}, this extends to bricks and rocks and exceeds our factor~$13/12$ for rocks with $\{0,1\}$-speeds (\Cref{thm:UB_Bricks_Rocks_13/12}). So general speeds are strictly harder than~$\{0,1\}$-speeds.

\subsection{Balanced and cyclic bags: a tight $25/16$ guarantee}
\label{sec:speeds-cyclic}

We next present a common algorithm for sand, bricks and rocks. For sand, we distribute the volume equally among the~$M$ bags, so each bag has volume $1/M$.

For bricks and rocks, order the jobs in Phase~I so that $p_1\ge\cdots\ge p_n$ and assign them cyclically to~$B_1,\ldots,B_M$. That is,
\[
 B_r=\{r,r+M,r+2M,\ldots\}\cap[n],
 \qquad r\in[M].
\]
Assigning one of these bags to each of $M$ identical machines and processing its jobs in~SPT order gives an optimal schedule. The same holds after restricting the construction to any prefix of the jobs in this order. In particular, for every prefix the job counts of bags differ by at most one. This property will allow us to
transfer the guarantee for bricks to rocks using~\Cref{lem:prefix-transfer}. This order of the jobs is used only to construct the bags.

In Phase~II, let $x_i$ denote the number of bags already assigned to machine~$i$. We initialize $x_i=0$ and assign all~$M$ bag labels $B_1,\ldots,B_M$ in this order, including those of empty bags. For each bag, choose an available machine minimizing the marginal value
\[
 \frac{(x_i+1)^2-x_i^2}{s_i}
 =\frac{2x_i+1}{s_i}
\]
and increase $x_i$ by one. We call this the \emph{quadratic marginal assignment}. This rule minimizes the increase in~$\sum_i x_i^2/s_i$, which is proportional to the cost for balanced sand bags. For bricks and rocks, we use the same rule which depends only on bag counts and machine speeds. Once the bag-to-machine assignment is fixed, we process jobs on each machine in SPT order. We call the resulting algorithms the \emph{cyclic-bag algorithm} for bricks and rocks and the \emph{balanced-bag algorithm} for sand.

\begin{restatable}{theorem}{restateSpeedsUB}
\label{thm:speeds-upper}\label{thm:speeds-cyclicbags-2516}
For general speeds, the balanced-bag algorithm for sand and the cyclic-bag  algorithm for bricks and rocks are $25/16$-robust for minimizing the total completion time. This bound is tight uniformly over the number of machines for these bag constructions, even with an optimal Phase~II assignment.
\end{restatable}
\begin{proof}[Upper bound for sand] 
Let $x_i$ be the final number of bags on machine~$i$. Since every bag has volume~$1/M$ and the total speed is one, the algorithm's ratio is
\[
 R(s)=\frac1{M^2}\sum_{i=1}^m\frac{x_i^2}{s_i}.
\]
Let $\lambda>0$ denote the marginal value selected when assigning the last bag~$B_M$. The marginal values on each machine increase, so the values selected by the algorithm are non-decreasing. If $x_i>0$, the last selected value on machine~$i$ is $(2x_i-1)/s_i$, while its next unselected value is~$(2x_i+1)/s_i$. Hence,
\[
 \frac{2x_i-1}{s_i}\le\lambda\le\frac{2x_i+1}{s_i}.
\]
Thus, $|2x_i-\lambda s_i|\le1$ and $\lambda s_i\ge2x_i-1\ge1$. If $x_i=0$, the first marginal value remains unselected, so~$0<\lambda s_i\le1$. In either case, $(2x_i-\lambda s_i)^2\le\lambda s_i$. Rearranging gives
\[
 \frac{x_i^2}{s_i}
 \le
 \lambda x_i+\frac{\lambda}{4}
 -\frac{\lambda^2s_i}{4}.
\]
Summing over all available machines, including those receiving no bags, and using $\sum_i x_i=M$ and $\sum_i s_i=1$, we obtain
\begin{align*}
 R(s)
 &\le \frac{\lambda(4M+m-\lambda)}{4M^2}
 = \frac{(4M+m)^2-(2\lambda-4M-m)^2}{16M^2}
 \le \frac{(4M+m)^2}{16M^2}
 \le \frac{25}{16},
\end{align*}
where the last inequality follows from $m\le M$.\qed
\end{proof}

\begin{proof}[sketch for upper bound for bricks and rocks]
For bricks, we must also account for the linear term in $F(k)=k(k+1)/2$ and the possible difference in bag sizes. Write~$n=qM+\ell$, where $0\le\ell<M$. The first $\ell$ bags contain $q+1$ jobs and the remaining bags contain $q$ jobs. Let $u_i$ and $x_i$ denote the bag counts on machine~$i$ after $\ell$ and $M$ assignments, respectively. Machine~$i$ then processes $k_i=qx_i+u_i$ jobs. Both count vectors satisfy the threshold inequalities used for sand, with their respective thresholds. We define the threshold for zero steps to be zero.

The algorithm and the optimum may assign different numbers of jobs to a machine, so we cannot compare their costs separately on each machine. Instead, we allow correction terms whose total over all machines is non-positive. Let~$z_i$ be the number of jobs on machine~$i$ in an unrestricted optimal schedule. Using the two sets of threshold inequalities, we prove in~\Cref{app:speeds-cyclic} that there are nonnegative values $P,G$, common to all machines, such that
\[
 \frac{F(k_i)}{s_i}
 \le
 \frac{25}{16}\frac{F(z_i)}{s_i}
 +P(k_i-z_i)+G(1-x_i).
\]
After summing over all machines, the job-count corrections cancel because both schedules process $n$ jobs. The bag count corrections sum to $G(m-M)\le0$, giving the desired bound. For rocks, both phases of the algorithm are consistent on prefixes of the job sequence, so \Cref{lem:prefix-transfer} applies as for \Cref{thm:UB_Bricks_Rocks_13/12}. \qed
\end{proof}

Tightness is shown in \Cref{app:speeds-tight}: with one fast machine of speed $3M/(5M-2)$ and $M-1$ equally slow ones, even an optimal assignment of balanced bags puts about~$3/4$ of the volume on the fast machine, although its share of the speed tends to $3/5$.

\subsection{Imbalanced bags for sand}
The tightness of $25/16$ for balanced bags (\Cref{thm:speeds-cyclicbags-2516}) shows that improving the guarantee requires a different Phase~I construction. With balanced bags, every machine that is used receives a volume of at least $1/M$. Smaller bags allow us to assign less volume to slow machines, while larger bags carry the remaining volume. For sand, we use this flexibility to obtain an improved guarantee.

In Phase~I, let $h=\lfloor M/2\rfloor$. We create $h$ bags of volume $3/(4M)$ and $h$ bags of volume $5/(4M)$. If $M$ is odd, we add one intermediate bag of volume $1/M$. In either case, there are $M$ bags of total volume one. Let~$b_1,\ldots,b_M$ denote their volumes.

In Phase~II, we first determine the number of bags assigned to each machine, independently of the bag volumes. Starting with $x_i = 0$, we repeat the following $M$ times: choose an available machine minimizing
\[
 \frac{3x_i+1}{s_i}
\]
and increase $x_i$ by one. The selection score for the first bag on machine~$i$ is still $1/s_i$, while each assigned bag increases the next score by $3/s_i$ instead of $2/s_i$. Compared with the quadratic marginal assignment, this makes choosing an unused machine relatively cheaper than adding another bag to an occupied machine.

After fixing the counts, we determine the explicit bag-to-machine assignment. We order the machines with $x_i>0$ in non-increasing order of $x_i/s_i$ and the bags by non-decreasing volume. In this order, each machine receives the $x_i$ smallest remaining bags. Thus, machines with less speed per assigned bag receive smaller bags, whereas machines with more speed per assigned bag receive larger bags. We call the resulting algorithm the \emph{imbalanced-bag algorithm}.

\begin{restatable}{theorem}{restateSpeedsImbalanced}
\label{thm:sand}\label{thm:speeds-imbalanced-sand}
For general speeds, the imbalanced-bag algorithm is $1681/1152$-robust for minimizing the total completion time of sand, where $1681/1152\approx1.4592$.
\end{restatable}

\begin{proof}[sketch]
The bag volumes satisfy $M\sum_r b_r^2\le17/16$, so their total squared volume is at most $17/16$ times that of balanced bags. To obtain an improved guarantee, we also account for how the small and large bags are assigned. By Cauchy--Schwarz, the contribution of machine~$i$ to the ratio $R(s)$ satisfies
\[
\frac{\left(\sum_{r:\pi(r)=i}b_r\right)^2}{s_i}
 \le
 \frac{x_i}{s_i}\sum_{r:\pi(r)=i}b_r^2.
\]
For the chosen counts, our assignment minimizes the sum of these upper bounds by pairing the smaller squared bag volumes with the larger coefficients $x_i/s_i$.

Let $\lambda$ be the largest score selected when fixing the bag counts. For a bag~$r$ assigned to machine~$i$, set~$t_r=\max\{1/3,x_i/(\lambda s_i)\}$. The count rule gives $t_r\in[1/3,1]$. Indexing the bags by non-decreasing volume, our assignment gives $t_1\ge\cdots\ge t_M$. Using the Cauchy--Schwarz bound and the threshold inequalities for all available machines, including those with $x_i=0$, we obtain
\begin{align*}
 R(s)
 &\le \lambda\sum_{r=1}^M t_rb_r^2 \le
 \left(\sum_{r=1}^M
       \left(\frac{3}{2t_r}-\frac12\right)\right)
 \left(\sum_{r=1}^M t_rb_r^2\right).
\end{align*}
The first inequality follows from Cauchy--Schwarz and the definition of $t_r$. The threshold inequalities bound~$\lambda$ by the first factor in the product.
 
We maximize this upper bound over all ordered coefficients in $[1/3,1]$. For fixed bag volumes, the product is convex when equal coefficients are varied together while preserving their order. Hence, its maximum over ordered coefficients is attained when the first $k$ coefficients equal $1$ and the remaining coefficients equal $1/3$.

The resulting bound depends on the sum of squared volumes of all bags and the corresponding sum for the $k$ smallest bags. Substituting the bag volumes gives two quadratic expressions in $k/M$, according to whether $k\le M/2$ or $k\ge M/2$. Both are bounded by $1681/1152$. For odd $M$, the intermediate bag preserves the same estimates on these sums of squared volumes.
\qed
\end{proof}

\begin{credits}
\subsubsection{\ackname}
This work was supported by the Deutsche Forschungsgemeinschaft (DFG, German Research Foundation) under Germany's Excellence Strategy -- EXC 3036 The Martian Mindset -- project number~533607631.

\subsubsection{AI Disclosure}
The results for identical machines in \Cref{sec:Equal_Speeds} were developed primarily by the authors. The authors developed the sand lower bound for general speeds in \Cref{thm:speeds-sand-lower}, the balanced-bag upper bound for sand in \Cref{thm:speeds-cyclicbags-2516}, and the prefix transfer in \Cref{lem:prefix-transfer}. Their construction for identical machines motivated the imbalanced bags in \Cref{thm:speeds-imbalanced-sand}. They also developed the non-adaptive results in \Cref{app:non-adap-detail} and, for the strong-adversary variants in \Cref{app:remarks-proofs}, the impossibility result for general speeds, the indivisible variant and its $9/8$ upper bound for sand, and the lower bound for bricks and rocks in \Cref{theorem:LB_Indisible_Strong_ADV}.

ChatGPT Astra generated initial proof ideas and assisted with checking and simplifying proofs, including those for identical machines. It generated the initial proof for bricks in \Cref{thm:speeds-cyclicbags-2516} and drafted the initial proof of \Cref{thm:speeds-imbalanced-sand} based on the authors’ imbalanced-bag idea. The algorithm and proof in \Cref{theorem:Opt_For_Strong_ADV} and the sand lower bound in \Cref{theorem:LB_Indisible_Strong_ADV} were also initially AI-generated. The authors substantially revised and simplified these proofs, including their technical arguments and, for bricks, the presentation of the scheduling intuition. ChatGPT Astra and Claude Opus 5.5 also assisted with manuscript review, presentation, and structure. The authors take full responsibility for the content and correctness of this work. They grant reviewers permission to upload this submission to generative AI services for the purpose of peer review.
\end{credits}

%
\bibliographystyle{splncs04}
\bibliography{references.bib}

\clearpage 
\appendix
\renewcommand*{\theHsection}{appendix.\Alph{section}}
\crefalias{section}{appendix}
\crefalias{subsection}{appendix}
\section*{Appendices}

\section{Transfer results (\Cref{sec:prelim})}
\label{app:prelim-proofs}

\subsection{From bricks to rocks (\Cref{lem:prefix-transfer})}
\restatePrefixTransfer*
\begin{proof}
Fix an instance $I$ with processing times $p_1\ge\cdots\ge p_n$
and a revealed speed vector $s$. Set $p_{n+1}=0$, and consider the assignment produced by the algorithm. For every $r \in [n]$, let $A_r$ denote the cost of restricting this assignment to the first $r$ jobs and setting their processing times to one. Similarly, let $\opt_r(s)$ be the unrestricted optimum for $r$ unit-size jobs and speed vector $s$. By assumption, $A_r \le \alpha \opt_r(s)$.

We now decompose the processing times in the original instance into layers. For every $r \in [n]$, the layer of height~$p_r-p_{r+1}$ contributes to precisely the first $r$ jobs. Whenever $p_r-p_{r+1}  > 0$, these jobs form a suffix of the SPT schedule on each machine. If machine $i$ contains $h$ of them, the layer contributes
\[
 \frac{p_r-p_{r+1}}{s_i}(1+\cdots+h)
 =(p_r-p_{r+1})\frac{h(h+1)}{2s_i}
\]
to its total completion time. Summing over all
machines and layers gives
\[
 \alg(I,s)
 =\sum_{r=1}^n(p_r-p_{r+1})A_r
 \le
 \alpha\sum_{r=1}^n(p_r-p_{r+1})\opt_r(s).
\]

We can apply the same decomposition to an unrestricted optimal SPT schedule for instance~$I$. For every $r$,
restricting its assignment to the first $r$ jobs
and setting their processing times to one gives
a feasible schedule of cost at least~$\opt_r(s)$.
Hence,
\[
 \opt(I,s)
 \ge
 \sum_{r=1}^n(p_r-p_{r+1})\opt_r(s).
\]
Combining the two inequalities yields
$\alg(I,s)\le\alpha\opt(I,s)$. \qed
\end{proof}

\subsection{From sand to bricks and rocks (\Cref{obs:sand-to-bricks})}
\restateSandToBricks*
\begin{proof}
Fix $M$, and let $\gamma$ be a lower bound on the robustness factor for sand. Consider any algorithm for bricks and an instance $I_n$ consisting of $n$ unit-size jobs. After Phase~I, interpret a bag containing $b$ jobs as a sand bag of volume $b/n$. These volumes sum to one.

Fix $\varepsilon>0$. By the lower bound for sand, there is an admissible speed vector~$s$ for which every assignment of these bags has ratio at least $\gamma-\varepsilon$. Write $S:=\sum_i s_i$. If the algorithm assigns $a_i$ jobs to machine~$i$, the corresponding sand load is $a_i/n$. Hence,
\[
 \alg(I_n,s)
 =\sum_{i:s_i>0}\frac{a_i(a_i+1)}{2s_i}
 \ge \sum_{i:s_i>0}\frac{a_i^2}{2s_i}
 \ge \frac{(\gamma-\varepsilon)n^2}{2S}.
\]

To bound the unrestricted optimum, assign each job independently to machine~$i$ with probability $s_i/S$. Let $X_i$ denote the resulting number of jobs on machine~$i$, and let $m\le M$ be the number of positive-speed machines. Since
\[
 \mathbb{E}[X_i(X_i+1)]
 =n(n-1)\left(\frac{s_i}{S}\right)^2
   +2n\frac{s_i}{S},
\]
the expected total completion time is
\[
 \sum_{i:s_i>0}
 \frac{\mathbb{E}[X_i(X_i+1)]}{2s_i}
 =\frac{n(n-1)}{2S}+\frac{nm}{S}
 \le \frac{n(n+2M-1)}{2S}.
\]
Some assignment has cost at most this expectation, so the same upper bound holds for $\opt(I_n,s)$. Consequently,
\[
 \frac{\alg(I_n,s)}{\opt(I_n,s)}
 \ge
 (\gamma-\varepsilon)\frac{n}{n+2M-1}.
\]
Letting $n$ tend to infinity and then $\varepsilon$ tend to zero proves the lower bound for bricks. The bound for rocks follows because bricks are a special case. \qed
\end{proof}

\subsection{From makespan to total completion time (\Cref{lemma:Makespan_To_MinSum})}
\restateMakespanToMinSum*

\begin{proof}
Let $S:=\sum_i s_i$, and let $y_i$ denote the volume
assigned to machine~$i$ by the algorithm. Since the
total volume is one, the optimal makespan is $1/S$.
The algorithm's makespan
\[
 T:=\max_{i:s_i>0}\frac{y_i}{s_i}
\]
therefore satisfies $T\le\gamma/S$. Using
$\opt(s)=1/(2S)$, we obtain
\[
 \frac{\alg(s)}{\opt(s)}
 =S\sum_{i:s_i>0}\frac{y_i^2}{s_i}
 \le ST\sum_{i:s_i>0}y_i
 =ST
 \le\gamma.
\]
\qed
\end{proof}

\section{Proofs for speeds in $\{0,1\}$ (\Cref{sec:Equal_Speeds})}\label{app:equal-speeds-proofs}
\subsection{Equal bags for sand (\Cref{theorem:Sand_Equally_UB})}\label{app:sand-equal}
\begin{observation}
\label[observation]{theorem:Sand_Equally_UB}
    For speeds in $\{0,1\}$, the algorithm that distributes sand equally among the bags and assigns them as evenly as possible in Phase~II is $9/8$-robust for minimizing the total completion time.
    This bound is tight for the algorithm.
\end{observation}
\begin{proof}
    Write $M=qm+r$ with $q\geq1$ and $0\leq r<m$. In Phase~II, each machine receives~$q$ bags, and $r$ machines receive one additional bag. Only these additional bags cause an imbalance between the machine loads since all bags have size $1/M$. Consequently,
    \[
     R(m)
     =\frac{m}{M^2}\bigl((m-r)q^2+r(q+1)^2\bigr)
     =1+\frac{r(m-r)}{M^2}.
    \]
    Since $M=mq+r\geq m+r$, we obtain
    \begin{align*}
        R(m) &\leq1+\frac{r(m-r)}{(m+r)^2}.
    \end{align*}
    Moreover, $(m+r)^2-8r(m-r)=(m-3r)^2\geq0$, and hence $R(m)\leq1+1/8=9/8$.
    
    Observe that for $M=4$ and $m=3$, the algorithm yields machine loads of $(1/2, 1/4, 1/4)$ and the ratio is $R(3)=\frac{3}{16}(2^2+1^2+1^2)=9/8$. Hence, the bound is tight.\qed
\end{proof}

\subsection{$3{:}2$ pattern for sand (\Cref{theorem:Sand_2_Sizes})}\label{app:sand-two-sizes}
We show that the robustness factor can be significantly improved even with the simple~$3{:}2$ pattern using two different bag sizes.
\begin{lemma}\label{theorem:Sand_2_Sizes}
    For speeds in $\{0,1\}$, consider the algorithm that distributes sand equally among the bags for~$M\leq2$ and, for $M\geq3$, uses two bag sizes in a ratio of $3{:}2$, with the larger size assigned to the first $\lfloor M/2\rfloor$ bags. In Phase~II, it assigns the bags in non-increasing order of volume to a least-loaded machine, breaking ties by machine index (LPT). This algorithm is $18/17$-robust for minimizing the total completion time, and this bound is tight for the algorithm.
\end{lemma}

\begin{proof}
For $M\leq2$, the algorithm is optimal. Let $M\geq3$, $L:=\lfloor M/2\rfloor$, and $T:=2M+L$. The algorithm uses~$L$ bags of volume $3/T$ and $M-L$ bags of volume $2/T$. For the analysis, multiply all volumes by $T$ and write $u_i:=Ty_i$ for the resulting integer machine loads. Thus,
\[
 \sum_{i=1}^m u_i=T
 \qquad\text{and}\qquad
 R(m)=\frac{m}{T^2}\sum_{i=1}^m u_i^2.
\]
For $m\leq L$, we bound the difference between machine loads. For $m>L$, the long bags occupy distinct machines, and we distinguish whether there are enough short bags to assign two to every remaining machine.

\medskip\noindent
\textbf{Case 1: $m\leq L$.}
After assigning the bags of size $3$, the machine loads are equal or differ by $3$. In the latter case, each least loaded machine receives one bag of size $2$ before any other machine does. There are enough such bags, since~$M-L\geq m$, and the load difference becomes $1$. Further assignments of bags of size $2$ preserve a load difference of at most $2$. The same bound holds if the loads were initially equal.

Let $a:=\min_i u_i$. Since the mean minimizes the sum of squared deviations,
\[
 \sum_{i=1}^m\left(u_i-\frac{T}{m}\right)^2
 \leq\sum_{i=1}^m(u_i-a-1)^2
 \leq m.
\]
Consequently,
\[
 R(m)
 =1+\frac{m}{T^2}
       \sum_{i=1}^m\left(u_i-\frac{T}{m}\right)^2
 \leq1+\frac{m^2}{T^2}
 \leq\frac{26}{25}<\frac{18}{17},
\]
where we used $m\leq L$ and $T\geq5L$.

\medskip\noindent
\textbf{Case 2: $L<m\leq(M+L)/2$.}
The bags of size $3$ are assigned to distinct machines. Since $2(m-L)\leq M-L$, LPT assigns two bags of size $2$ to each of the other $m-L$ machines. The remaining $M+L-2m$ bags are assigned to machines of load $3$. Thus, exactly $m-L$ machines have load $4$, and the remaining $L$ machines have load $3$ or $5$.
It follows that
\[
 \sum_{i=1}^m(u_i-4)^2=L,
 \qquad\text{and hence}\qquad
 \sum_{i=1}^m u_i^2=8T-16m+L.
\]
Since $L/T\leq1/5$, we obtain
\begin{align*}
 R(m)
 &=\left(8+\frac LT\right)\frac mT
   -16\left(\frac mT\right)^2\\
 &\leq\frac{41}{5}\frac mT
   -16\left(\frac mT\right)^2
 \leq\frac{1681}{1600}<\frac{18}{17},
\end{align*}
where we bounded the quadratic by completing the square.

\medskip\noindent
\textbf{Case 3: $(M+L)/2<m\leq M$.}
In this case, all long bags remain alone, while each remaining machine receives one or two short bags. Thus, exactly $L$ machines have load~$3$, and the remaining machines have load~$2$ or $4$.

First, suppose $m<M$. Then $m\leq M-1\leq2L$, so at least half the machines have load~$3$. Consequently,
\[
 \sum_{i=1}^m(u_i-3)^2=m-L\leq\frac m2,
\]
and expanding the squares gives
\[
 \sum_{i=1}^m u_i^2\leq6T-\frac{17}{2}m.
\]
It follows that
\begin{align*}
 R(m)
 &\leq6\frac mT-\frac{17}{2}\left(\frac mT\right)^2\\
 &=\frac{18}{17}
   -\frac{17}{2}\left(\frac mT-\frac6{17}\right)^2
 \leq\frac{18}{17}.
\end{align*}
For $m=M$, every bag is assigned separately, and
\[
 R(M)=1+\frac{L(M-L)}{T^2}
 \leq\frac{25}{24}<\frac{18}{17},
\]
where the inequality follows from
\[
 T^2-24L(M-L)=(2M-5L)^2\geq0.
\]

For tightness, take $M=7$ and $m=6$. The bag volumes are $(3,3,3,2,2,2,2)/17$, and LPT produces machine loads~$(4,3,3,3,2,2)/17$.
Thus,
\[
 R(6)
 =\frac6{17^2}\left(4^2+3\cdot3^2+2\cdot2^2\right)
 =\frac{18}{17}.
\]
\qed
\end{proof}

The lemma also determines the asymptotic behavior of the algorithm from \Cref{sec:Equal_Speeds} for bricks.
\begin{remark}\label{rem:bricks-asymptotic}
    For every fixed $M\geq3$, the normalized bag loads converge to the two-size distribution of \Cref{theorem:Sand_2_Sizes}, and, on this limiting distribution, the Phase~II assignment corresponds to LPT up to tie-breaking. Since $F(k)=k^2/2+O(k)$, \Cref{alg:ALG_Bricks_13/12} is asymptotically $18/17$-robust as $n\to\infty$, which is tight for $M=7$ and $m=6$.
\end{remark}

\subsection{Linear bag profiles for sand (\Cref{theorem:Sand_UB_Improved})}
\restateSandLinear*
\begin{proof}
Let $x(t):=2(1+t)/3$. In Phase~I, we define the volume of bag~$B_i$ by
\[
 b_i:=\int_{(i-1)/M}^{i/M}x(t)\,dt
     =\frac{2}{3M}+\frac{2i-1}{3M^2},
 \qquad i\in[M].
\]
Thus, the bag volumes form an increasing arithmetic progression and sum to one.

We first reduce the analysis to $M/2<m\leq M$. If $m\leq M/2$, let $h$ be the smallest power of two with~$hm>M/2$. Then $hm\leq M$, since $(h/2)m\leq M/2$. Phase~II first assigns the bags to $hm$ virtual machines using the pairing rule below, and then merges consecutive groups $G_1,\ldots,G_m$ of $h$ virtual machines each. If their loads are $z_1,\ldots,z_{hm}$, Cauchy--Schwarz gives
\[
 R(m)
 =m\sum_{j=1}^m\left(\sum_{\ell\in G_j}z_\ell\right)^2
 \leq mh\sum_{\ell=1}^{hm}z_\ell^2
 =R(hm).
\]
Thus, merging these groups cannot increase the ratio.

Now let $M/2<m\leq M$, and put $k:=M-m$ and $s:=2k/M\in[0,1)$. Phase~II pairs the $2k$ smallest bags in opposite order and assigns each pair and each remaining bag to a separate machine. There are $k$ pairs and $M-2k$ remaining bags, hence exactly~$m$ machines. All pairs have the same volume, since
\[
 b_i+b_{2k+1-i}
 =\frac{4(M+k)}{3M^2}
 =\frac{2(2+s)}{3M},
 \qquad 1\leq i\leq k.
\]

It remains to bound the contribution of the bags assigned alone. For each such bag, let $I_i=[(i-1)/M,i/M]$. The average of $x$ over $I_i$ is $Mb_i$. Jensen's inequality for $z\mapsto z^2$ bounds its square by the average of $x^2$:
\[
 (Mb_i)^2
 =\left(\frac{1}{|I_i|}\int_{I_i}x(t)\,dt\right)^2
 \leq\frac{1}{|I_i|}\int_{I_i}x(t)^2\,dt
 =M\int_{I_i}x(t)^2\,dt.
\]
Dividing by $M^2$ and summing over the remaining bags, whose intervals partition $[s,1]$, gives
\[
 \sum_{i=2k+1}^M b_i^2
 \leq\frac1M\int_s^1x(t)^2\,dt
 =\frac{4}{9M}\int_s^1(1+t)^2\,dt.
\]
Using $k/M=s/2$ and $m/M=1-s/2$, we obtain
\begin{align}
 R(m)
 &=m\left(
       k\left(\frac{2(2+s)}{3M}\right)^2
       +\sum_{i=2k+1}^M b_i^2
     \right)\notag\\
 &\leq\frac49\left(1-\frac{s}{2}\right)
       \left(
         \frac{s}{2}(2+s)^2+\int_s^1(1+t)^2\,dt
       \right)\notag\\
 &=\frac49\left(1-\frac{s}{2}\right)
       \left(\frac73+s+s^2+\frac{s^3}{6}\right)
 =:H(s).\label{eq:Sand_Linear_H}
\end{align}
It remains to bound $H(s)$ independently of $s$. Expanding and regrouping gives
\[
 H(s)=\frac{28}{27}
 -\frac{s}{27}\left(2(1-s)(1-2s)^2+s(3s-2)^2\right)
 \leq\frac{28}{27},
\]
because every term in parentheses is nonnegative for $s\in[0,1]$. Together with the merging argument, this proves the bound for every $m\in[M]$.

For $m=M$, assigning each bag to a separate machine is optimal and gives $R(M)=28/27-1/(27M^2)$. This tends to $28/27$ as $M\to\infty$, proving that the factor is tight uniformly over $M$ for these bag volumes. \qed
\end{proof}

\subsection{Lower bound for sand (\Cref{theorem:Sand_LB_1_033})}
\restateSandLB*

\begin{proof}
We prove the lower bound by choosing a fixed distribution over the number of available machines and showing that every bag construction has an expected ratio of at least $31/30$. Consequently, at least one machine count gives the claimed lower bound.

Fix $M=12$ and any bag volumes $b_1\geq\cdots\geq b_{12}\geq0$ with $\sum_{i=1}^{12}b_i=1$. Let $R(m)$ denote the smallest ratio achievable by assigning these bags to $m$ machines. We choose the number of available machines according to the distribution 
\[
 \Pr[m=8]=\frac{15}{45},\quad
 \Pr[m=9]=\frac4{45},\quad
 \Pr[m=10]=\frac6{45},\quad
 \Pr[m=12]=\frac{20}{45}.
\]
This distribution is independent of the bag volumes.

We first establish two properties of an optimal assignment. For $m\in\{8,9,10\}$, we may assume that every machine receives at least one bag, possibly of volume zero. Indeed, moving a bag from a machine containing at least two bags to an empty machine cannot increase the sum of squared loads. At least $v:=2m-12$ machines then receive exactly one bag.

Moreover, we may assume that the $v$ largest bags are assigned alone. Suppose that a singleton bag has volume $b$, while a bag of volume $a\geq b$ shares its machine with other bags of total volume $c$. Exchanging the two bags changes the sum of squared loads by
\[
 a^2+(b+c)^2-b^2-(a+c)^2=2c(b-a)\leq0.
\]
After setting aside the $v$ largest bags and their machines, the remaining bags occupy $m-v=12-m$ machines. Applying Cauchy--Schwarz to these machines gives
\[
 R(m)\geq m\left(
   \sum_{i=1}^v b_i^2
   +\frac{1}{12-m}\left(\sum_{i=v+1}^{12}b_i\right)^2
 \right).
\]

For $m=8,9,10$, the values of $v$ are $4,6,8$. We therefore partition the bags into the corresponding consecutive groups and write
\[
 X_1:=\sum_{i=1}^4b_i,\qquad
 X_2:=b_5+b_6,\qquad
 X_3:=b_7+b_8,\qquad
 X_4:=\sum_{i=9}^{12}b_i.
\]
Their volumes sum to one. For $m=12$, assigning every bag to a separate machine is optimal. Applying Cauchy--Schwarz within the groups yields
\begin{align*}
 R(8)&\geq2\bigl(X_1^2+(X_2+X_3+X_4)^2\bigr),\\
 R(9)&\geq\frac94X_1^2+\frac92X_2^2+3(X_3+X_4)^2,\\
 R(10)&\geq\frac52X_1^2+5(X_2^2+X_3^2+X_4^2),\\
 R(12)&\geq3X_1^2+6X_2^2+6X_3^2+3X_4^2.
\end{align*}

These bounds capture the different effects of the machine counts. With fewer machines, more bags must share machines, and their combined volume contributes quadratically. With more machines, the individual bag volumes determine more of the cost. Averaging the four bounds according to the chosen distribution gives
\begin{align*}
 \mathbb{E}[R(m)]
 &\geq\frac{38}{15}X_1^2+\frac23(1-X_1)^2\\
 &\quad+\frac{56}{15}X_2^2+\frac{18}{5}X_3^2+\frac{34}{15}X_4^2+\frac8{15}X_3X_4.
\end{align*}
We first combine $X_3$ and $X_4$ into their total volume, and then combine this volume with $X_2$. This leaves a quadratic expression in $X_1$ and $1-X_1$, which we can bound independently of the bag volumes. For this, we use the weighted form of Cauchy--Schwarz: for $\alpha,\beta>0$,
\[
 \alpha u^2+\beta v^2\geq\frac{(u+v)^2}{1/\alpha+1/\beta}.
\]
Let $Z:=X_3+X_4$. Completing a square gives
\begin{align*}
 \frac{18}{5}X_3^2+\frac{34}{15}X_4^2+\frac8{15}X_3X_4
 &=\frac1{12}(5X_3-3X_4)^2+\frac{91}{60}Z^2\\
 &\geq\frac{91}{60}Z^2.
\end{align*}
Since $X_2+Z=1-X_1$, the weighted inequality implies
\begin{align*}
    \frac{56}{15}X_2^2+\frac{91}{60}Z^2
    &\geq\frac{(X_2+Z)^2}{\frac{15}{56}+\frac{60}{91}}\\
    &=\frac{728}{675}(1-X_1)^2.
\end{align*}
Consequently,
\begin{align*}
 \mathbb{E}[R(m)]
 &\geq\frac{38}{15}X_1^2+\left(\frac23+\frac{728}{675}\right)(1-X_1)^2\\
 &=\frac{38}{15}\left(X_1^2+\frac{31}{45}(1-X_1)^2\right)\\
 &\geq\frac{38}{15}\cdot\frac{(X_1+(1-X_1))^2}{1+45/31}\\
 &=\frac{38}{15}\cdot\frac{31}{76}\\
 &=\frac{31}{30}.
\end{align*}
Thus, for every bag construction, the expected ratio under the chosen distribution is at least $31/30$. At least one of the four machine counts therefore forces this ratio, even with an optimal Phase~II assignment.
\qed
\end{proof}

\subsection{$3{:}2$-Cyclic algorithm for bricks and rocks
(\Cref{thm:UB_Bricks_Rocks_13/12})}\label{app:Bricks_App}

\begin{algorithm}[tb]
    \caption{$3{:}2$-Cyclic}
    \label{alg:ALG_Bricks_13/12}
    \begin{algorithmic}[1]
        \Require Phase I: $n$ jobs, $M$ bags
        \Require Phase II: Number of machines $m$

        \State Sort the jobs in non-increasing order of processing times.
        \Statex \textbf{Case $M\leq 2$:}
        \State \textit{Phase I:} Assign the jobs cyclically to the bags.
        \State \textit{Phase II:} Assign one bag to each machine if $m=2$, and all bags to the single machine otherwise. Schedule the jobs in SPT order.
        
        \Statex \textbf{Case $M\geq 3$:}
        \State Set $L:=\lfloor M/2\rfloor$ and $x:=2M+L$.

        \Statex \textit{Phase I:}
        \State Partition the given job sequence into consecutive packages of $x$ jobs, with possibly fewer jobs in the last package.
        \State For each package, assign its jobs cyclically to $B_1,\ldots,B_M$, starting with $B_1$.
        
        \Statex \textit{Phase II:}
        \State Assign $B_1,\ldots,B_L$ cyclically to machines $1,\ldots,m$, starting with machine~1.
        \State Set $h:=m\lceil L/m\rceil-L$.
        \If{$h>0$}
            \State Assign $B_{L+1},\ldots,B_{L+h}$ to machines $m-h+1,\ldots,m$, respectively.
        \EndIf
        \State Assign the remaining bags in increasing index order, cyclically to machines $m,m-1,\ldots,1$.
        \State Schedule the jobs in SPT order on each machine.
    \end{algorithmic}
\end{algorithm}

\restateBricksRocksUB*

\begin{figure}[hbt]
    \centering
    \scalebox{0.6}{
      \begin{tikzpicture}[
      x=1cm,y=1cm,
      block/.style   = {draw,minimum width=3,anchor=south west},
      hatch/.style   = {block,pattern=north east lines,pattern color=violet,draw=violet},
      brace/.style   = {decorate,decoration={brace,mirror,amplitude=4pt}}
        ]
        \node[font=\Large] at (-3.4,4.5) {Phase I};
        \node[font=\Large] at (7.2,4.5) {Phase II};
        \draw[line width=1.5pt] (-0.0, -1.35) -- (-0.0,5.25);
        
        \foreach \i [evaluate=\i as \j using int(\i+1)] in {0,...,6}{
            \filldraw[draw=black, opacity=1-0.035*\i, line width=0.2pt, fill=job1]
            (-6.3+0.85*\i,0.0) -- (-6.3+0.85*\i,0.5) -- (-5.6+0.85*\i,0.5) -- (-5.6+0.85*\i,0.0) -- cycle;
            \node[font=\Large] at (-5.95+0.85*\i,0.25) {\j};
            \node[font=\Large] at (-5.95+0.85*\i,-0.4) {$B_{\j}$};
        }
        \foreach \i [evaluate=\i as \j using int(\i+8)] in {0,...,6}{
            \filldraw[draw=black, opacity=1-0.035*(7+\i), line width=0.2pt, fill=job1]
            (-6.3+0.85*\i,0.6) -- (-6.3+0.85*\i,1.1) -- (-5.6+0.85*\i,1.1) -- (-5.6+0.85*\i,0.6) -- cycle;
            \node[font=\Large] at (-5.95+0.85*\i,0.85) {\j};
        }
        \foreach \i [evaluate=\i as \j using int(\i+15)] in {0,...,2}{
            \filldraw[draw=black, opacity=1-0.035*(13+\i), line width=0.2pt, fill=job1]
            (-6.3+0.85*\i,1.2) -- (-6.3+0.85*\i,1.7) -- (-5.6+0.85*\i,1.7) -- (-5.6+0.85*\i,1.2) -- cycle;
            \node[font=\Large] at (-5.95+0.85*\i,1.45) {\j};
        }
        \foreach \i [evaluate=\i as \j using int(\i+18)] in {0,...,2}{
            \filldraw[draw=black, opacity=1-0.035*(17+\i), line width=0.2pt, fill=job2]
            (-6.3+0.85*\i,1.8) -- (-6.3+0.85*\i,2.3) -- (-5.6+0.85*\i,2.3) -- (-5.6+0.85*\i,1.8) -- cycle;
            \node[font=\Large] at (-5.95+0.85*\i,2.05) {\j};
        }
        \filldraw[draw=black, opacity=1-0.035*20, line width=0.2pt, fill=job2]
            (-6.3+0.85*3,1.2) -- (-6.3+0.85*3,1.7) -- (-5.6+0.85*3,1.7) -- (-5.6+0.85*3,1.2) -- cycle;
            \node[font=\Large] at (-5.95+0.85*3,1.45) {21};
        
        \node[font=\Large] at (0.80+0.9*1, -1.2) {$m=3$};
        \node[font=\Large] at (0.80+0.9*0,-0.5) {$m_{1}$};
        \node[font=\Large] at (0.80+0.9*1,-0.5) {$m_{2}$};
        \node[font=\Large] at (0.80+0.9*2,-0.5) {$m_{3}$};
        
        \filldraw[draw=black, opacity=1-0.075*(0), line width=0.2pt, fill=job4] (0.45+0.9*0,0.0) -- (0.45+0.9*0,2.0) -- (1.15+0.9*0,2.0) -- (1.15+0.9*0,0.0) -- cycle;
        \node[font=\Large] at (0.80+0.9*0,1.0) {$B_1$};
        \filldraw[draw=black, opacity=1-0.075*(1), line width=0.2pt, fill=job4] (0.45+0.9*1,0.0) -- (0.45+0.9*1,2.0) -- (1.15+0.9*1,2.0) -- (1.15+0.9*1,0.0) -- cycle;
        \node[font=\Large] at (0.80+0.9*1,1.0) {$B_2$};
        \filldraw[draw=black, opacity=1-0.075*(2), line width=0.2pt, fill=job4] (0.45+0.9*2,0.0) -- (0.45+0.9*2,2.0) -- (1.15+0.9*2,2.0) -- (1.15+0.9*2,0.0) -- cycle;
        \node[font=\Large] at (0.80+0.9*2,1.0) {$B_3$};
        \filldraw[draw=black, opacity=1-0.075*(3), line width=0.2pt, fill=job3] (0.45+0.9*2,2.1) -- (0.45+0.9*2,3.7) -- (1.15+0.9*2,3.7) -- (1.15+0.9*2,2.1) -- cycle;
        \node[font=\Large] at (0.80+0.9*2,2.9) {$B_4$};
        \filldraw[draw=black, opacity=1-0.075*(4), line width=0.2pt, fill=job3] (0.45+0.9*1,2.1) -- (0.45+0.9*1,3.3) -- (1.15+0.9*1,3.3) -- (1.15+0.9*1,2.1) -- cycle;
        \node[font=\Large] at (0.80+0.9*1,2.7) {$B_5$};
        \filldraw[draw=black, opacity=1-0.075*(5), line width=0.2pt, fill=job3] (0.45+0.9*0,2.1) -- (0.45+0.9*0,3.3) -- (1.15+0.9*0,3.3) -- (1.15+0.9*0,2.1) -- cycle;
        \node[font=\Large] at (0.80+0.9*0,2.7) {$B_6$};
        \filldraw[draw=black, opacity=1-0.075*(6), line width=0.2pt, fill=job3] (0.45+0.9*2,3.8) -- (0.45+0.9*2,5.0) -- (1.15+0.9*2,5.0) -- (1.15+0.9*2,3.8) -- cycle;
        \node[font=\Large] at (0.80+0.9*2,4.40) {$B_7$};

        \node[font=\Large] at (4.35+0.9*1.5,-1.2) {$m=4$};
        \node[font=\Large] at (4.35+0.9*0,-0.5) {$m_{1}$};
        \node[font=\Large] at (4.35+0.9*1,-0.5) {$m_{2}$};
        \node[font=\Large] at (4.35+0.9*2,-0.5) {$m_{3}$};
        \node[font=\Large] at (4.35+0.9*3,-0.5) {$m_{4}$};
        
        \filldraw[draw=black, opacity=1-0.075*(0), line width=0.2pt, fill=job4] (4.0+0.9*0,0.0) -- (4.0+0.9*0,2.0) -- (4.7+0.9*0,2.0) -- (4.7+0.9*0,0.0) -- cycle;
        \node[font=\Large] at (4.35+0.9*0,1.0) {$B_1$};
        \filldraw[draw=black, opacity=1-0.075*(1), line width=0.2pt, fill=job4] (4.0+0.9*1,0.0) -- (4.0+0.9*1,2.0) -- (4.7+0.9*1,2.0) -- (4.7+0.9*1,0.0) -- cycle;
        \node[font=\Large] at (4.35+0.9*1,1.0) {$B_2$};
        \filldraw[draw=black, opacity=1-0.075*(2), line width=0.2pt, fill=job4] (4.0+0.9*2,0.0) -- (4.0+0.9*2,2.0) -- (4.7+0.9*2,2.0) -- (4.7+0.9*2,0.0) -- cycle;
        \node[font=\Large] at (4.35+0.9*2,1.0) {$B_3$};
        \filldraw[draw=black, opacity=1-0.075*(3), line width=0.2pt, fill=job3] (4.0+0.9*3,0.0) -- (4.0+0.9*3,1.6) -- (4.7+0.9*3,1.6) -- (4.7+0.9*3,0.0) -- cycle;
        \node[font=\Large] at (4.35+0.9*3,0.80) {$B_4$};
        \filldraw[draw=black, opacity=1-0.075*(4), line width=0.2pt, fill=job3] (4.0+0.9*3,1.7) -- (4.0+0.9*3,2.9) -- (4.7+0.9*3,2.9) -- (4.7+0.9*3,1.7) -- cycle;
        \node[font=\Large] at (4.35+0.9*3,2.30) {$B_5$};
        \filldraw[draw=black, opacity=1-0.075*(5), line width=0.2pt, fill=job3] (4.0+0.9*2,2.1) -- (4.0+0.9*2,3.3) -- (4.7+0.9*2,3.3) -- (4.7+0.9*2,2.1) -- cycle;
        \node[font=\Large] at (4.35+0.9*2,2.70) {$B_6$};
        \filldraw[draw=black, opacity=1-0.075*(6), line width=0.2pt, fill=job3] (4.0+0.9*1,2.1) -- (4.0+0.9*1,3.3) -- (4.7+0.9*1,3.3) -- (4.7+0.9*1,2.1) -- cycle;
        \node[font=\Large] at (4.35+0.9*1,2.70) {$B_7$};

        \node[font=\Large] at (8.80+0.9*2.5,-1.2) {$m=6$};
        \node[font=\Large] at (8.80+0.9*0,-0.5) {$m_{1}$};
        \node[font=\Large] at (8.80+0.9*1,-0.5) {$m_{2}$};
        \node[font=\Large] at (8.80+0.9*2,-0.5) {$m_{3}$};
        \node[font=\Large] at (8.80+0.9*3,-0.5) {$m_{4}$};
        \node[font=\Large] at (8.80+0.9*4,-0.5) {$m_{5}$};
        \node[font=\Large] at (8.80+0.9*5,-0.5) {$m_{6}$};
        
        \filldraw[draw=black, opacity=1-0.075*(0), line width=0.2pt, fill=job4] (8.45+0.9*0,0.0) -- (8.45+0.9*0,2.0) -- (9.15+0.9*0,2.0) -- (9.15+0.9*0,0.0) -- cycle;
        \node[font=\Large] at (8.80+0.9*0,1.0) {$B_1$};
        \filldraw[draw=black, opacity=1-0.075*(1), line width=0.2pt, fill=job4] (8.45+0.9*1,0.0) -- (8.45+0.9*1,2.0) -- (9.15+0.9*1,2.0) -- (9.15+0.9*1,0.0) -- cycle;
        \node[font=\Large] at (8.80+0.9*1,1.0) {$B_2$};
        \filldraw[draw=black, opacity=1-0.075*(2), line width=0.2pt, fill=job4] (8.45+0.9*2,0.0) -- (8.45+0.9*2,2.0) -- (9.15+0.9*2,2.0) -- (9.15+0.9*2,0.0) -- cycle;
        \node[font=\Large] at (8.80+0.9*2,1.0) {$B_3$};
        \filldraw[draw=black, opacity=1-0.075*(3), line width=0.2pt, fill=job3] (8.45+0.9*3,0.0) -- (8.45+0.9*3,1.6) -- (9.15+0.9*3,1.6) -- (9.15+0.9*3,0.0) -- cycle;
        \node[font=\Large] at (8.80+0.9*3,0.80) {$B_4$};
        \filldraw[draw=black, opacity=1-0.075*(4), line width=0.2pt, fill=job3] (8.45+0.9*4,0.0) -- (8.45+0.9*4,1.2) -- (9.15+0.9*4,1.2) -- (9.15+0.9*4,0.0) -- cycle;
        \node[font=\Large] at (8.80+0.9*4,0.60) {$B_5$};
        \filldraw[draw=black, opacity=1-0.075*(5), line width=0.2pt, fill=job3] (8.45+0.9*5,0.0) -- (8.45+0.9*5,1.2) -- (9.15+0.9*5,1.2) -- (9.15+0.9*5,0.0) -- cycle;
        \node[font=\Large] at (8.80+0.9*5,0.60) {$B_6$};
        \filldraw[draw=black, opacity=1-0.075*(6), line width=0.2pt, fill=job3] (8.45+0.9*5,1.3) -- (8.45+0.9*5,2.5) -- (9.15+0.9*5,2.5) -- (9.15+0.9*5,1.3) -- cycle;
        \node[font=\Large] at (8.80+0.9*5,1.90) {$B_7$};
        \end{tikzpicture}
        }
      \caption{Illustration of both phases for $n=21$ bricks and $M=7$ bags.
      Left: Phase~I with $L=3$, $x=17$, and an incomplete second package (orange), resulting in unequal short bags.
      Right: Phase~II for three machine numbers, with long bags (pink) and short bags (green).}
      \label{fig:Example_Phase_I_and_II}
\end{figure}
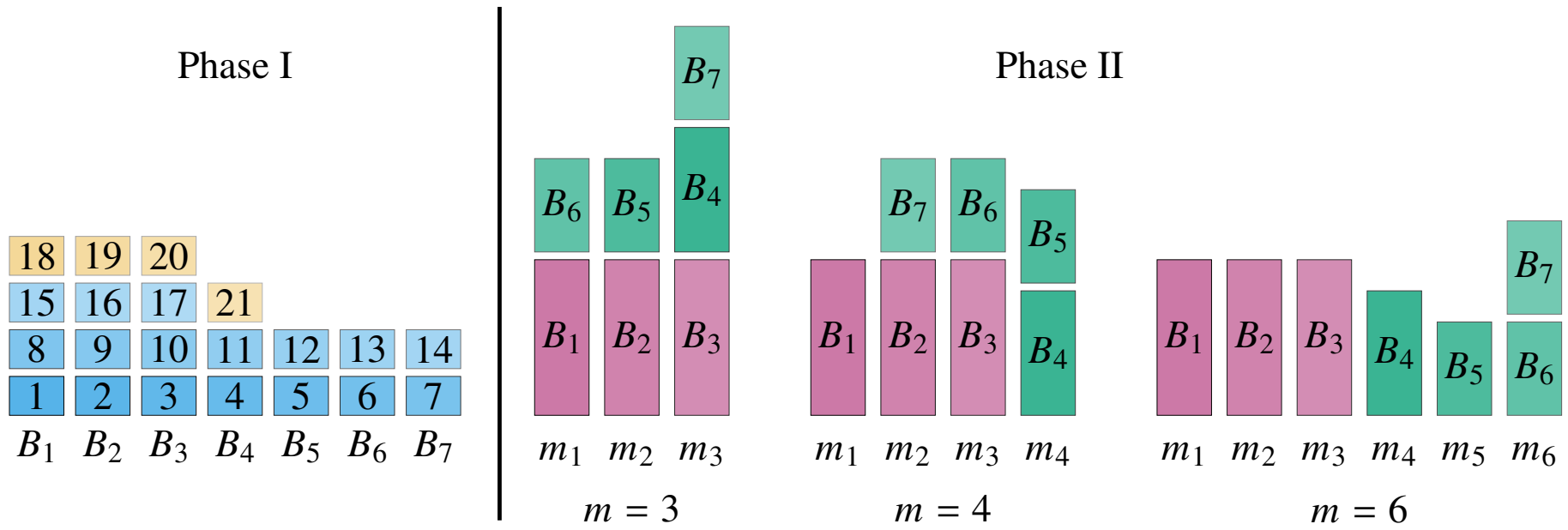

\Cref{alg:ALG_Bricks_13/12} gives a formal description of the algorithm from \Cref{sec:Equal_Speeds}, and \Cref{fig:Example_Phase_I_and_II} illustrates both phases. In Phase~II, we first assign bags in complete forward cyclic rounds until all long bags have been assigned. We then reverse the cyclic order for the remaining bags. The assignment applies to all bag labels, including those of empty bags, and depends only on $M$ and $m$.

We first prove the guarantee for bricks and then extend it to rocks using \Cref{lem:prefix-transfer}. Let $M\geq3$, $L:=\lfloor M/2\rfloor$, and~$b_i:=|B_i|$ for $i\in[M]$. We refer to $B_1,\ldots,B_L$ as long bags and to the remaining bags as short bags. Recall that~$F(k)=k(k+1)/2$ gives the total completion time of $k$ bricks on one machine.

By construction, the bag loads satisfy
\begin{equation}\label{eq:cyclic-bag-shape}
\begin{aligned}
 b_1&\geq\cdots\geq b_M,\\
 b_{L+1}-b_M&\leq1,\\
 \left|b_i-\frac32b_j\right|&\leq1
 \qquad\text{for all }i\leq L<j.
\end{aligned}
\end{equation}
The first two inequalities follow from the cyclic assignment within each package. For the third inequality, each complete package adds three jobs to every long bag and two jobs to every short bag. Thus, these packages do not change $b_i-\frac32b_j$. The contributions of the last, possibly incomplete package to a long bag and a short bag form one of the pairs
\[
 (0,0),\ (1,0),\ (1,1),\ (2,1),\ (2,2),\ (3,2),
\]
where the first entry refers to the long bag. Each pair satisfies the claimed inequality.

We next bound the difference between machine loads. The forward rounds give non-increasing machine loads, whereas the reverse rounds add non-decreasing loads. We use these opposite orders to bound the final difference between machine loads.

\begin{lemma}\label{lemma:Max_Machine_Load_Diff}
Let $s:=b_{L+1}$.
The machine loads produced by the algorithm differ by at most $\max\{1,s\}$.
\end{lemma}
\begin{proof}
Write $a:=b_M$. By \eqref{eq:cyclic-bag-shape}, we have
\[
 b_1-b_M\leq\frac a2+1.
\]
Since the bag loads are integers, this implies
\[
 b_1-b_M\leq\max\{1,a\}\leq\max\{1,s\}.
\]
Recall that $K=m\lceil L/m\rceil$ is the number of bags assigned before the algorithm reverses the cyclic order. Then~$L\leq K\leq M$. Indeed, if $m\leq L$, then $K<L+m\leq2L\leq M$. Otherwise, $K=m\leq M$.

Let $y_i^0$ denote the load assigned to machine~$i$ in the forward rounds. Since the bags are ordered by non-increasing load, we have $y_1^0\geq\cdots\geq y_m^0$ and
\[
 y_1^0-y_m^0
 =\sum_{r=0}^{K/m-1}
   \bigl(b_{rm+1}-b_{(r+1)m}\bigr)
 \leq b_1-b_K.
\]
The inequality follows by adding the nonnegative gaps between consecutive rounds.

Let $d_i$ denote the additional load assigned to machine~$i$ in the reverse rounds. These loads satisfy $d_1\leq\cdots\leq d_m$. Setting $b_j:=0$ for $j>M$, the same argument gives
\[
 d_m-d_1
 =\sum_{r\geq0}
   \bigl(b_{K+rm+1}-b_{K+(r+1)m}\bigr)
 \leq b_{K+1}.
\]
For $K=M$, both sides are zero.

The final load of machine~$i$ is $y_i=y_i^0+d_i$. Consequently, for $i<j$,
\[
 -b_{K+1}\leq y_i-y_j\leq b_1-b_K.
\]
The claim follows from $K\geq L$, $b_{K+1}\leq s$, and $b_1-b_K\leq b_1-b_M\leq\max\{1,s\}$.
\qed
\end{proof}

For bricks on identical machines, an unrestricted optimal schedule has machine loads differing by at most one. Indeed, moving one job from a machine with at least two more jobs than another machine strictly decreases the total completion time. The following lemma bounds the cost when the loads differ by two.

\begin{lemma}\label{lemma:Balance_ALG_to_Opt}
A schedule of bricks with machine loads in $\{\beta,\beta+1,\beta+2\}$, where $\beta\geq2$, has cost at most $13/12$ times the unrestricted optimum.
\end{lemma}
\begin{proof}
As long as there is a machine with load $\beta$ and a machine with load $\beta+2$, move one job from the latter to the former. Each such move decreases the total completion time by one. When no further move is possible, the loads differ by at most one, so the resulting schedule is optimal.

Let $k$ be the number of moves. Then the original cost exceeds the optimum by $k$. Moreover, the resulting schedule has at least $2k$ machines of load $\beta+1\geq3$, each contributing at least $F(3)=6$.
Thus, $\opt_m\geq12k$, and the original cost is at most
\[
 \opt_m+k\leq\frac{13}{12}\opt_m.
\]
\qed
\end{proof}

We now prove the robustness guarantee. If no short bag is assigned alone, every machine receives a long bag or at least two short bags. This gives a lower bound on its load compared with the possible difference between machine loads. If a short bag is assigned alone, all long bags are assigned alone as well. Their loads are close to the midpoint between the loads of one short bag and two short bags. We use these properties to bound the variance in the two cases.

\begin{proof}[Proof of \Cref{thm:UB_Bricks_Rocks_13/12}]
We first consider bricks. For $M\leq2$, the algorithm produces machine loads differing by at most one and is therefore optimal. Hence, assume $M\geq3$ and let $s:=b_{L+1}$.

If $s\leq1$, the machine loads differ by at most one by \Cref{lemma:Max_Machine_Load_Diff}, so the algorithm is optimal. Suppose that~$s=2$. By~\eqref{eq:cyclic-bag-shape}, every short bag contains at least one job and every long bag contains at least two jobs. If every machine has load at least two, the claim follows from \Cref{lemma:Max_Machine_Load_Diff,lemma:Balance_ALG_to_Opt}.

Otherwise, some machine has load one and therefore receives a short bag alone. This requires~$m>L$, since for~$m\leq L$ every machine receives a long bag. For~$m>L$, the first $z:=2m-M$ bags are assigned alone, and all remaining bags are assigned in pairs. Since one of the first $z$ bags has load one and all bags are nonempty, we have~$b_z=1$. Thus, all bags assigned in pairs have load one. Moreover, \eqref{eq:cyclic-bag-shape}, applied to a short bag of load one, implies that every long bag has load at most two. All machine loads therefore belong to $\{1,2\}$, and the algorithm is optimal.

Henceforth, assume $s\geq3$. Let $y_i$ denote the load of machine~$i$, and write
\[
 \overline y:=\frac nm,
 \qquad
 V:=\frac1m\sum_{i=1}^m(y_i-\overline y)^2.
\]
Since $F(k)=k(k+1)/2$, we have
\[
 \alg(m,M)
 =\frac m2\bigl(V+\overline y^2+\overline y\bigr).
\]
By convexity,
\[
 \opt_m\geq\frac m2
 \bigl(\overline y^2+\overline y\bigr).
\]
Consequently, 
\begin{equation}\label{eq:cyclic-variance-ratio}
 \frac{\alg(m,M)}{\opt_m}
 \leq
 1+\frac{V}{\overline y^2+\overline y}.
\end{equation}
For the remaining cases, we either apply \Cref{lemma:Balance_ALG_to_Opt} or show that $12V\leq\overline y^2+\overline y$.

In each case, we will locate the machine loads in an interval and choose its midpoint~$c$. We bound their squared distances from $c$ and then account for the difference between $c$ and the average load. Let
\[
 D_c:=\frac1m\sum_{i=1}^m(y_i-c)^2,
 \qquad
 \tau:=\overline y-c.
\]
Then $V=D_c-\tau^2$. Completing squares gives a lower bound on $\overline y^2+\overline y-12V$ that depends only on $c$ and $D_c$:
\begin{equation}\label{eq:cyclic-reference-load}
\begin{aligned}
 \overline y^2+\overline y-12V
 &=
 12\left(\tau+\frac c{12}\right)^2
 +\left(\tau+\frac12\right)^2
 +\frac{11}{12}c^2+c-12D_c-\frac14\\
 &\geq
 \frac{11}{12}c^2+c-12D_c-\frac14.
\end{aligned}
\end{equation}
We distinguish whether a short bag is assigned alone.

\medskip\noindent
\textbf{Case 1: No short bag is assigned alone.} Every machine receives a long bag or at least two short bags. By~\eqref{eq:cyclic-bag-shape}, every long bag has load at least $3s/2-1$, and every short bag has load at least $s-1$. Thus, the minimum machine load $\beta$ satisfies
\[
 \beta
 \geq\min\left\{\frac32s-1,\,2(s-1)\right\}
 =\frac32s-1,
\]
where the equality uses $s\geq3$.

By \Cref{lemma:Max_Machine_Load_Diff}, all machine loads belong to $[\beta,\beta+s]$. Choose
\[
 c:=\beta+\frac s2\geq2s-1.
\]
Then every machine load differs from $c$ by at most $s/2$, so $D_c\leq s^2/4$. Applying \eqref{eq:cyclic-reference-load}, we obtain
\[
\begin{aligned}
 \overline y^2+\overline y-12V
 &\geq
 \frac{11}{12}(2s-1)^2+(2s-1)-3s^2-\frac14\\
 &=\frac{2s^2-5s-1}{3}>0,
\end{aligned}
\]
where the last inequality holds for $s\geq3$. The claim follows from \eqref{eq:cyclic-variance-ratio}.

\medskip\noindent
\textbf{Case 2: A short bag is assigned alone.} Then $m>L$ and $z:=2m-M>L$. The first~$z$ bags are assigned alone, and the remaining bags are assigned in pairs. In particular, all~$L$ long bags are assigned alone.

Let $a:=b_z$. Since $B_z$ is a short bag, we have $a\geq s-1\geq2$. The short bags assigned alone have loads in~$[a,a+1]$. All bags assigned in pairs are short and have loads in~$[a-1,a]$. Hence, every machine receiving two bags has load in
\[
 [2a-2,2a]\subseteq[a,2a].
\]
By \eqref{eq:cyclic-bag-shape}, the loads of the long bags belong to
\[
 \left[\frac32a-1,\frac32a+1\right]
 \subseteq[a,2a].
\]

If $a=2$, all machine loads belong to $\{2,3,4\}$, and \Cref{lemma:Balance_ALG_to_Opt} proves the claim. Thus, assume $a\geq3$ and choose $c:=3a/2$. The loads of the $L$ machines receiving long bags differ from $c$ by at most one. Every other machine has load in $[a,2a]$ and therefore differs from $c$ by at most $a/2$. Since
\[
 \frac Lm\geq\frac LM\geq\frac13
 \qquad\text{and}\qquad
 \frac{a^2}{4}\geq1,
\]
we obtain
\[
 D_c
 \leq
 \frac Lm+
 \left(1-\frac Lm\right)\frac{a^2}{4}
 \leq\frac13+\frac{a^2}{6}.
\]
Applying \eqref{eq:cyclic-reference-load} now gives
\[
\begin{aligned}
 \overline y^2+\overline y-12V
 &\geq
 \frac{11}{12}\left(\frac32a\right)^2
 +\frac32a
 -12\left(\frac13+\frac{a^2}{6}\right)-\frac14\\
 &=\frac{a^2}{16}+\frac32a-\frac{17}{4}>0,
\end{aligned}
\]
where the last inequality holds for $a\geq3$. Again, \eqref{eq:cyclic-variance-ratio} proves the claim.

Finally, consider rocks with $p_1\geq\cdots\geq p_n$. In Phase~I, the assignment of jobs to bag labels depends only on their indices and $M$. In Phase~II, the assignment of all bag labels, including empty bags, depends only on~$M$ and $m$. Thus, for every $r\in[n]$, restricting the assignment to the first $r$ jobs and setting their processing times to one gives the algorithm's assignment for $r$ bricks. The guarantee therefore holds for every prefix under the same assignment. Applying \Cref{lem:prefix-transfer} completes the proof. \qed
\end{proof}

\subsection{Lower bound for rocks (\Cref{thm:LB_Bricks})}\label{app:rocks-lb}
\restateBricksLB*
\begin{proof}
The case $n=6$ follows from the lower bound for bricks. For $n\geq7$, fix $M:=3$, let $k:=n-6$, and add $k$ jobs of size $\varepsilon\in(0,1)$ to the six unit jobs.

Consider any partition of this instance into three bags. Restricting the bags to the unit jobs gives the same case distinction as in \Cref{sec:Equal_Speeds}. If each bag contains two unit jobs, the adversary chooses $m=2$, and the unit jobs contribute at least $13$. Otherwise, the adversary chooses $m=3$, and their contribution is at least~$10$. The additional jobs cannot decrease these contributions.

To bound the unrestricted optimum, start with an optimal schedule of the six unit jobs. Its cost is $12$ for~$m=2$ and $9$ for $m=3$. Place all additional jobs first on one machine. Their own completion times and the delay of at most three unit jobs increase the cost by at most
\[
 \delta:=\frac{k(k+1)}2\varepsilon+3k\varepsilon
 =\frac{k(k+7)}2\varepsilon.
\]
Consequently,
\[
 \frac{\alg}{\opt}
 \geq\min\left\{
   \frac{13}{12+\delta},
   \frac{10}{9+\delta}
 \right\}.
\]
For every fixed $n\geq7$, we have $\delta\to0$ as $\varepsilon\to0$, and the right-hand side tends to $13/12$. Since this bound holds for every bag partition, no algorithm can guarantee a robustness factor smaller than $13/12$.
\qed
\end{proof}

\section{Proofs for general speeds (\Cref{sec:speeds})}
\label{app:speeds-proofs}

\subsection{Lower bound for sand (\Cref{thm:speeds-sand-lower})}
\restateSpeedsSandLB*
\begin{proof}
Fix $M\ge2$ and arbitrary bag volumes $0\le b_1\le\cdots\le b_M$ summing to one. Let~$W(b)$ denote the supremum, over all normalized speed vectors, of the ratio achieved by an optimal Phase~II assignment of these bags. Recall that an assignment with machine loads $y_i$ has ratio
\[
 R(s)=\sum_{i=1}^m\frac{y_i^2}{s_i},
\]
since both the total volume and the total speed are one.

Set $\alpha = (M-1)/M$. We first describe the minimum relaxed quadratic cost of selecting a prescribed total volume to be assigned to slow machines. Define $q:[0,1]\to\mathbb{R}_{\ge 0}$ by linearly interpolating points
\[
 \left(
   \sum_{r=1}^k b_r,\,
   (M-1)\sum_{r=1}^k b_r^2
 \right),
 \qquad k=0,\ldots,M.
\]
For a prescribed volume of $u$, consider the relaxation
\[
 \min\left\{
   (M-1)\sum_{r=1}^M z_rb_r^2:
   \sum_{r=1}^M z_rb_r=u,\quad
   0\le z_r\le1
 \right\}.
\]
The minimum is obtained by selecting the smallest bags first, with at most one bag selected fractionally. Hence, its value is exactly $q(u)$.

The segment corresponding to a positive bag volume $b_r$ has length $b_r$ and slope $(M-1)b_r$. Thus, $q$ is convex, $q(0)=0$, and 
\begin{equation}
\label{eq:sand-lb-reciprocal}
 \int_0^1\frac{du}{q'(u)}
 =
 \frac{|\{r:b_r>0\}|}{M-1}
 \le\frac1\alpha.
\end{equation}
Here and below, derivatives of piecewise differentiable functions are understood almost everywhere and bags with a volume of zero only contribute duplicate interpolation points.

We now relate this profile to a speed vector that forces a large ratio. More precisely, we show that 
\begin{equation}
\label{eq:sand-lb-h}
 W(b)\ge h^2,
 \qquad
 h:=\max_{0\le u\le1}
       \left\{\sqrt{q(u)}+1-u\right\}.
\end{equation}
For $0 <\sigma\le \alpha$, give $M-1$ machines speed $\sigma/(M-1)$ and the remaining machine a speed of $1-\sigma$. Consider any assignment, and let $T\subseteq[M]$ be the indices of the bags assigned to the slow machines. Dropping the cross terms between bags assigned to the same slow machine gives the lower bound
\[
 R(s)\ge
 \frac{M-1}{\sigma}\sum_{r\in T}b_r^2
 +\frac{\left(1-\sum_{r\in T}b_r\right)^2}{1-\sigma}.
\]
By relaxing the choice of $T$ as above, we obtain
\begin{equation}
\label{eq:sand-lb-relaxation}
 W(b)\ge
 \min_{0\le u\le1}
 \left\{
   \frac{q(u)}{\sigma}
   +\frac{(1-u)^2}{1-\sigma}
 \right\}.
\end{equation}

For uniform speeds, assigning one bag to each machine is optimal, since combining bags only adds nonnegative cross terms. The resulting ratio is
\[
 U=M\sum_{r=1}^M b_r^2=\frac{q(1)}{\alpha}.
\]
If $U \ge h^2$, then \eqref{eq:sand-lb-h} follows immediately. Otherwise, a maximizing point $u_*$ in the definition of $h$ lies in $(0,1)$: the squared values of the two endpoints are $1$ and $q(1)$, both of which are at most $U$. Write
\[
 A=\sqrt{q(u_*)},\qquad B=1-u_*,
 \qquad h=A+B.
\]
In particular, $A,B>0$. By definition of $h$, the graph of $q$ lies below the parabola $(h-1+u)^2$, with equality at~$u_*$. Since $q$ lies below the parabola and agrees with it at~$u_*$, we have $q'_-(u_*)\ge2A$ and $q'_+(u_*)\le2A$. Convexity gives~$q'_-(u_*)\le q'_+(u_*)$, so both derivatives equal $2A$. The corresponding tangent lies below $q$, so
\[
 q(u)\ge A^2+2A(u-u_*)
 \qquad\text{for all }u\in[0,1].
\]
In particular,
$q(1)\ge A^2+2AB\ge Ah$. Hence, the choice
\[
 \sigma=\frac Ah
 \le\frac{q(1)}{h^2}<\alpha
\]
is admissible. Substituting the tangent bound gives
\[
 \frac{q(u)}{\sigma}
 +\frac{(1-u)^2}{1-\sigma}
 \ge
 h^2+\frac hB(u-u_*)^2
 \ge h^2
 \qquad\text{for all }u\in[0,1].
\]
Together with \eqref{eq:sand-lb-relaxation}, this proves \eqref{eq:sand-lb-h}.

We next use the restriction to $M$ bags to derive a lower bound on $h$. The first segment of $q$ has positive slope, so~$\sqrt{q(u)}>u$ for sufficiently small $u>0$, and hence $h>1$. If~$h \ge 2$, the desired bound is immediate. We therefore assume that $a:=h-1$ lies in~$(0,1)$. The definition of $h$ gives
\[
 q(u)\le(u+a)^2
 \qquad\text{for all }u\in[0,1].
\]
Moreover, convexity and $q(0)=0$ imply
\[
 q(u)\le\frac ua q(a)\le4au
 \qquad\text{for }0\le u\le a.
\]
Consequently, $q(u)\le v(u)$, where
\[
 v(u)=
 \begin{cases}
  4au, & 0\le u\le a,\\
  (u+a)^2, & a\le u\le1.
 \end{cases}
\]
The function $v$ follows the line from the origin tangent to the parabola at $u=a$, and then follows the parabola itself. In particular, $v'$ is positive, continuous and non-decreasing.

We now claim that
\[
 \int_0^1\frac{du}{q'(u)}
 \ge
 \int_0^1\frac{du}{v'(u)}.
\]
To see this, set $D=v-q\ge0$. Convexity of $z\mapsto1/z$ gives, almost everywhere,
\[
 \frac1{q'}-\frac1{v'}
 \ge
 \frac{v'-q'}{(v')^2}
 =
 \frac{D'}{(v')^2}.
\]
Integrating by parts and using the definition of $v$ gives
\[
 \int_0^1
 \left(\frac1{q'(u)}-\frac1{v'(u)}\right)\,du
 \ge
 \frac{D(1)}{4(1+a)^2}
 +\int_a^1\frac{D(u)}{2(u+a)^3}\,du
 \ge0,
\]
since $D\ge0$.

Combining the comparison with \eqref{eq:sand-lb-reciprocal}, we obtain
\[
 \frac1\alpha
 \ge
 \int_0^a\frac{du}{4a}
 +\int_a^1\frac{du}{2(u+a)}
 =
 \frac14+\frac12\log\frac{1+a}{2a}.
\]
Substituting $a=h-1$ and rearranging gives
\[
 h\ge
 \left(1-\frac12e^{\,1/2-2/\alpha}\right)^{-1}.
\]
By \eqref{eq:sand-lb-h}, every partition into $M$ bags therefore satisfies
\[
 W(b)\ge
 \left(
   1-\frac12
   \exp\left(\frac12-\frac{2M}{M-1}\right)
 \right)^{-2}.
\]
Letting $M\to\infty$ shows that no robustness factor valid for every number of machines can be smaller than
\[
 \gamma=
 \left(1-\frac12e^{-3/2}\right)^{-2}
 =1.2669187\ldots.
\]
\qed
\end{proof}

\subsection{Balanced and cyclic bags (\Cref{thm:speeds-cyclicbags-2516})}
\label{app:speeds-cyclic}
\restateSpeedsUB*

The statement for sand was proved in \Cref{sec:speeds-cyclic}. We first prove the upper bound for bricks and then extend it to rocks using \Cref{lem:prefix-transfer}. The tightness part is shown in \Cref{app:speeds-tight}.

\begin{proof}
Fix a speed vector $s$ with $m\leq M$ available machines and $\sum_{i=1}^m s_i=1$. Throughout this proof, sums over machines range over $i\in[m]$. Recall that the quadratic marginal assignment considers all $M$ bag labels in the order~$B_1,\ldots,B_M$, including empty bags. At each step, it chooses a machine minimizing $(2x_i+1)/s_i$, breaking ties according to the fixed machine order.

Let $x_i(h)$ denote the number of bags assigned to machine~$i$ after $h$ steps, and let $\lambda_h$ be the value selected in step~$h$. Set $\lambda_0=0$. Since the marginal values on each machine increase, the selected values are non-decreasing. Comparing the last selected and the next unselected value on each machine gives
\begin{equation}
\label{eq:cyclic-upper-thresholds}
 \max\{0,2x_i(h)-1\}
 \leq \lambda_h s_i
 \leq 2x_i(h)+1
 \qquad
 \text{for all }i\in[m],\ 0\leq h\leq M.
\end{equation}
If $x_i(h)=0$, the lower bound is immediate and the upper bound follows because the first value $1/s_i$ remains unselected. These inequalities also hold when $h=0$ and in the presence of ties.

\medskip\noindent\textit{Bricks.}
Write $n=qM+\ell$, where $0\leq\ell<M$. Each complete round of the cyclic construction contributes one job to every bag, while the last round contributes one additional job to the first $\ell$ bags. Set
\[
 x_i:=x_i(M),\qquad
 u_i:=x_i(\ell),\qquad
 \lambda:=\lambda_M,\qquad
 \mu:=\lambda_\ell.
\]

The complete rounds contribute $qx_i$ jobs to machine~$i$, and the last round contributes~$u_i$. Thus, its job count is~$k_i=qx_i+u_i$. Recall that its cost is $F(k_i)/s_i$, where $F(k)=k(k+1)/2$.

Let $z_i$ denote the number of jobs assigned to machine~$i$ in an unrestricted optimal schedule. The optimum may leave a machine unused even when the algorithm assigns jobs to it, so a separate multiplicative comparison on each machine is not possible. We therefore seek common nonnegative values $P,G$ such that
\begin{equation}
\label{eq:cyclic-machine-comparison}
 \frac{F(k_i)}{s_i}
 \leq
 \frac{25}{16}\frac{F(z_i)}{s_i}
 +P(k_i-z_i)+G(1-x_i).
\end{equation}

The terms $P(k_i-z_i)$ compensate for different job counts and cancel when summed over all machines. The remaining correction gives an allowance $G$ to each available machine and subtracts $G$ for each assigned bag. Its total is $G(m-M)\leq0$.

Suppose first that $q\geq1$. Apply \Cref{lem:cyclic-scalar-comparison} below with
\[
 t=\lambda s_i,\qquad
 w=\mu s_i,\qquad
 x=x_i,\qquad
 u=u_i,\qquad
 z=z_i.
\]
Its assumptions follow from \eqref{eq:cyclic-upper-thresholds}, $u_i\leq x_i$, and $\mu\leq\lambda$. Dividing the resulting inequality by $s_i$ gives \eqref{eq:cyclic-machine-comparison} with
\[
 P=\frac{(5q+3)\lambda+5\mu}{8},
 \qquad
 G=\frac{q(q-1)\lambda+2q\mu}{8}.
\]
Both values are nonnegative.

If $n<M$, every nonempty bag contains one job, so $q=0$ and $k_i=u_i$. We apply the same lemma with
\[
 (q,x,u,t,w,z)=(1,u_i,0,\mu s_i,0,z_i).
\]
This gives \eqref{eq:cyclic-machine-comparison} with $P=\mu$ and $G=0$.

Summing \eqref{eq:cyclic-machine-comparison} over all available machines yields
\begin{align*}
 \alg(s)
 &\leq
 \frac{25}{16}\opt(s)
 +P\sum_{i=1}^m(k_i-z_i)
 +G\sum_{i=1}^m(1-x_i)\\
 &=
 \frac{25}{16}\opt(s)+G(m-M)\\
 &\leq
 \frac{25}{16}\opt(s),
\end{align*}
since $\sum_i k_i=\sum_i z_i=n$ and $\sum_i x_i=M$.

\medskip\noindent\textit{Rocks.}
Fix an instance with $p_1\geq\cdots\geq p_n$. The assignment of jobs to bag labels depends only on their indices and $M$. The assignment map $\pi$ for all $M$ bag labels depends only on $M$ and $s$, including the fixed rule for breaking ties. Thus, for every $r\in[n]$, restricting the assignment to the first $r$ jobs and setting their processing times to one gives precisely the algorithm's assignment for $r$ bricks. The preceding argument therefore gives the guarantee for every prefix under the same assignment. Applying \Cref{lem:prefix-transfer} completes the proof.
\qed
\end{proof}

It remains to prove the comparison used for bricks. The following lemma combines the threshold inequalities for the complete rounds and the final partial round.

\begin{lemma}
\label{lem:cyclic-scalar-comparison}
Let $q\geq1$ and $0\leq u\leq x$ be integers. Suppose that $t,w\geq0$ satisfy
\[
 \max\{0,2x-1\}\leq t\leq2x+1,\qquad
 \max\{0,2u-1\}\leq w\leq2u+1,\qquad
 w\leq t.
\]
For $k=qx+u$ and every integer $z\geq0$, we have
\[
 F(k)
 \leq
 \frac{25}{16}F(z)
 +\frac{(5q+3)t+5w}{8}(k-z)
 +\frac{q(q-1)t+2qw}{8}(1-x).
\]
\end{lemma}

\begin{proof}
Set
\[
  \widehat P:=(5q+3)t+5w,
 \qquad
  \widehat G:=q(q-1)t+2qw,
\]
and write
\[
 H:=\frac{25}{16}F(z)-F(k)
       +\frac{\widehat P}{8}(k-z)+\frac{\widehat G}{8}(1-x).
\]
We show that $H\geq0$. We first reduce the threshold variables to endpoints and the counts to their smallest compatible values. We then minimize the resulting quadratic in $z$ and show that its minimum increases with the remaining parameters.

For fixed $q,x,u,z$, the expression $H$ is affine in $(t,w)$, so it suffices to consider vertices of their feasible region. If $u<x$, the constraint $w\leq t$ is redundant because $2u+1\leq2x-1$. If $u=x$, it cuts the corresponding square along its diagonal. Hence, every positive coordinate of a vertex is an odd integer. If~$t=0$, then~$w=x=u=0$ and~$H=25F(z)/16\geq0$. Thus, assume $t\geq1$.

Suppose first that $w>0$. Write
\[
 x=\frac{t-1}{2}+\alpha,
 \qquad
 u=\frac{w-1}{2}+\beta,
 \qquad
 \alpha,\beta\in\{0,1\}.
\]
Let $H_0$ be the value obtained by setting $\alpha=\beta=0$, with $k=qx+u$ adjusted accordingly. Direct subtraction gives
\begin{align*}
 8(H-H_0)
 &=
 q\alpha(4t-w)
 +\beta\bigl((q+3)t+w+4q-4\bigr)
 -8q\alpha\beta\\
 &\geq
 q\bigl(3\alpha+5\beta-8\alpha\beta\bigr)\\
 &=
 q\bigl(3\alpha(1-\beta)+5\beta(1-\alpha)\bigr)
 \geq0.
\end{align*}
Here, the first inequality follows from $t\geq w\geq1$.

If $w=0$, then $u=0$ and the two possible counts are $x=(t-1)/2$ and $x=(t+1)/2$. Increasing $x$ from the former to the latter increases $H$ by $q(t-1)/2\geq0$. Again, let~$H_0$ denote the value for the smaller count. Thus, in both cases it suffices to prove that~$H_0\geq0$.

We first consider $t=1$. Then $w\in\{0,1\}$ and both reduced counts are zero. Setting $v:=q+w$, we obtain
\[
 32H_0=(5z-2v+1)^2+3z-1.
\]
This expression is positive for integer $z\geq1$. For $z=0$, we have $H_0=v(v-1)/8\geq0$, since $v\geq1$.

It remains to consider $t\geq3$. Suppose first that $w\geq1$, and define
\[
 h(q,t,w):=\min_{z\in\mathbb R}H_0.
\]
Allowing real values of $z$ can only decrease the minimum. The quadratic is minimized at
\[
 z_*=\frac{4 \widehat P-25}{50}.
\]
After substituting this value, we show that the minimum increases with $q$ and $w$. Differentiating gives
\begin{align*}
 \frac{\partial h}{\partial q}
 &=
 \frac{20q(t-1)+8t^2-30t+25w}{80}\\
 &\geq
 \frac{8t^2-10t+5}{80}>0,\\
 \frac{\partial h}{\partial w}
 &=
 \frac{25q+3t+10w}{80}>0,
\end{align*}
where we used $q,w\geq1$ and $t\geq3$. Consequently,
\[
 h(q,t,w)\geq h(1,t,1).
\]
The remaining function also increases with $t\geq3$, since
\[
 \frac{d}{dt}h(1,t,1)=\frac{44t-85}{400}>0.
\]
Hence, it suffices to consider $q=1$, $t=3$, and $w=1$. At these values, the reduced counts are $x=1$ and~$u=0$, and
\[
 H_0=\frac{25z^2-91z+84}{32}.
\]
Its minimum is $119/3200>0$, so
\[
 H_0\geq h(q,t,w)\geq h(1,3,1)=\frac{119}{3200}>0.
\]

Finally, suppose that $w=0$. Let $h_0(q,t)$ denote the minimum of the corresponding expression $H_0$ over real $z$. The same quadratic minimization gives
\[
 h_0(q,t)-h(q,t,1)
 =
 \frac{(5q+12)t-5q-15}{80}
 \geq0,
\]
where the inequality uses $t\geq3$. The preceding argument therefore yields $H_0\geq h_0(q,t)\geq h(q,t,1)>0$.

Thus, $H_0\geq0$ in every case. Since $H\geq H_0$, the claim follows.
\qed
\end{proof}

\subsection{Tightness for balanced and cyclic bags (\Cref{thm:speeds-cyclicbags-2516})}
\label{app:speeds-tight}

We prove the tightness part of \Cref{thm:speeds-cyclicbags-2516}: for balanced bags for sand and cyclic bags for bricks and rocks, the robustness factor $25/16$ cannot be improved uniformly over the number of machines, even with an optimal Phase~II assignment.

\begin{proof}
Let $M\ge2$ and set $q=5M-2$. Consider $n=qM$ bricks and the speed vector
\[
 s_1=\frac{3M}{2},
 \qquad
 s_2=\cdots=s_M=1.
\]
Every cyclic bag contains exactly $q$ jobs. We use unnormalized speeds, since scaling all speeds does not change the ratio. The choice of $q$ will allow us to compare the bag assignment with an integral assignment of individual jobs in proportion to the speeds.

Consider the assignment of any Phase~II algorithm, and let $x_i$ be the number of bags assigned to machine~$i$. Its total completion time is
\[
 \alg(s)
 =
 \sum_{i=1}^M\frac{qx_i(qx_i+1)}{2s_i}
 \ge
 \frac{q^2}{2}\sum_{i=1}^M\frac{x_i^2}{s_i}.
\]
Since the bag counts are nonnegative integers, $x_i^2\ge x_i$. Using $\sum_i x_i=M$, we obtain
\begin{align*}
 \sum_{i=1}^M\frac{x_i^2}{s_i}
 &\ge
 \frac{2x_1^2}{3M}+\sum_{i=2}^M x_i\\
 &=
 \frac{2x_1^2}{3M}+M-x_1\\
 &=
 \frac{5M}{8}
 +\frac{2}{3M}
  \left(x_1-\frac{3M}{4}\right)^2\\
 &\ge\frac{5M}{8}.
\end{align*}
Thus, every assignment of the bags has cost at least $5q^2M/16$.

To upper-bound the unrestricted optimum, assign $z_i=2Ms_i$ jobs to machine~$i$. These numbers are integers: the fast machine receives $3M^2$ jobs, and each slow machine receives $2M$ jobs. Moreover, since~$\sum_i s_i=q/2$, this assignment contains exactly $Mq=n$ jobs. Its cost is
\[
 \sum_{i=1}^M\frac{z_i(z_i+1)}{2s_i}
 =
 2M^2\sum_{i=1}^M s_i+M^2
 =
 M^2(q+1).
\]
Consequently,
\[
 \frac{\alg(s)}{\opt(s)}
 \ge
 \frac{5q^2}{16M(q+1)}
 =
 \frac{5(5M-2)^2}{16M(5M-1)}
 \longrightarrow\frac{25}{16}
 \qquad\text{as }M\to\infty.
\]
Thus, the bound holds already for bricks and for every Phase~II assignment, including an optimal one.

The same speed vector proves tightness for balanced sand bags. Its total speed is~$S=(5M-2)/2$, and every assignment of bags of volume $1/M$ satisfies
\begin{align*}
    R(s)=
 \frac{S}{M^2}\sum_{i=1}^M\frac{x_i^2}{s_i}
 &\ge
 \frac{5(5M-2)}{16M}\\
 &=
 \frac{25}{16}-\frac{5}{8M}
 \longrightarrow\frac{25}{16}.
\end{align*}
\qed
\end{proof}

\subsection{Imbalanced bags for sand (\Cref{thm:speeds-imbalanced-sand})}
\restateSpeedsImbalanced*

\begin{proof}
Fix a speed vector $s$ with $m$ available machines and $\sum_{i=1}^m s_i=1$. Let $b_1\le\cdots\le b_M$ be the bag volumes in assignment order, let $\pi$ be the resulting bag-to-machine assignment, and let $x_i$ be the number of bags assigned to machine~$i$.

Let $\lambda$ be the largest selection score used when determining the bag counts. On each occupied machine, the last selected score is at most $\lambda$, and the next score is at least $\lambda$. Thus,
\[
 3x_i-2\le\lambda s_i\le3x_i+1
 \quad\text{if }x_i>0,
 \qquad
 \lambda s_i\le1
 \quad\text{if }x_i=0.
\]
For each bag $r\in[M]$, define
\[
 t_r=
 \max\left\{
   \frac13,\,
   \frac{x_{\pi(r)}}{\lambda s_{\pi(r)}}
 \right\},
 \qquad
 g(t)=\frac{3}{2t}-\frac12.
\]
The threshold inequalities imply $t_r\in[1/3,1]$. Moreover, the algorithm assigns bags in non-decreasing volume order to machines in non-increasing $x_i/s_i$ order. Since the definition of $t_r$ preserves this ordering, we have
\[
 1\ge t_1\ge\cdots\ge t_M\ge\frac13.
\]

We first relate $\lambda$ to these coefficients. All bags assigned to an occupied machine~$i$ have the same coefficient, and
\[
 \sum_{r:\pi(r)=i}g(t_r)
 =
 \min\left\{
   4x_i,\,
   \frac{3\lambda s_i-x_i}{2}
 \right\}
 \ge\lambda s_i+x_i-1.
\]
Indeed, the first expression inside the minimum satisfies the inequality by~$\lambda s_i\le3x_i+1$, and the second by~$\lambda s_i\ge3x_i-2$. For an unused available machine, the same inequality holds because its left-hand side is zero and $\lambda s_i\le1$. Summing over all available machines therefore gives
\begin{equation}
\label{eq:imbalanced-threshold-sum}
 \sum_{r=1}^M g(t_r)
 \ge
 \lambda\sum_{i=1}^m s_i+\sum_{i=1}^m x_i-m
 =
 \lambda+M-m
 \ge\lambda.
\end{equation}

We next bound the ratio of the assignment. By Cauchy--Schwarz on each occupied machine,
\begin{align*}
 R(s)
 &=
 \sum_{i:x_i>0}
 \frac{\left(\sum_{r:\pi(r)=i}b_r\right)^2}{s_i}\\
 &\le
 \sum_{i:x_i>0}
 \frac{x_i}{s_i}
 \sum_{r:\pi(r)=i}b_r^2\\
 &\le
 \lambda\sum_{r=1}^M t_rb_r^2.
\end{align*}
The last inequality follows from the definition of $t_r$. Together with \eqref{eq:imbalanced-threshold-sum}, this yields
\begin{equation}
\label{eq:imbalanced-product}
 R(s)
 \le
 \left(\sum_{r=1}^M g(t_r)\right)
 \left(\sum_{r=1}^M t_rb_r^2\right).
\end{equation}

It remains to bound the product on the right-hand side. We maximize it over all ordered coefficients $1\ge t_1\ge\cdots\ge t_M\ge1/3$, keeping the bag volumes fixed. This relaxation can only increase the upper bound. We show that it suffices to consider coefficients taking only the values $1/3$ and $1$.

Choose a value $v\in(1/3,1)$ occurring among the coefficients, and let
\[
 G=\{r:t_r=v\},\qquad q=|G|.
\]
Since the coefficients are ordered, $G$ consists of consecutive indices. Replace their common value by $u$, keeping all other coefficients fixed. To preserve the ordering, $u$ may vary between the nearest smaller and larger coefficient values, using the boundary $1/3$ or $1$ if no such value exists.

Define
\[
 A=\sum_{r\notin G}g(t_r),\qquad
 B=\sum_{r\notin G}t_rb_r^2,\qquad
 W=\sum_{r\in G}b_r^2.
\]
Thus, $A$ and $B$ are the fixed contributions to the two factors from indices outside $G$. The indices in $G$ contribute $qg(u)$ to the first factor and $uW$ to the second. Hence, the product becomes
\[
 \Phi(u)
 =
 (A+qg(u))(B+uW)
 =
 \left(A+\frac{3q}{2u}-\frac q2\right)(B+uW).
\]
The only nonlinear term depending on $u$ is $3qB/(2u)$. Since $B\ge0$, the function $\Phi$ is convex. We may therefore move~$u$ to an endpoint of its admissible interval without decreasing the product.

Such a move either merges $G$ with a neighboring group of equal coefficients or gives its coordinates the value $1/3$ or $1$. In either case, one distinct value in $(1/3,1)$ disappears. Repeating this step leaves, for some~$k\in\{0,\ldots,M\}$, the first $k$ coefficients equal to $1$ and the remaining $M-k$ coefficients equal to $1/3$.

Set
\[
 \theta=\frac{k}{M},\qquad
 Q=\sum_{r=1}^M b_r^2,\qquad
 Q_k=\sum_{r=1}^k b_r^2.
\]
Since $g(1)=1$ and $g(1/3)=4$, the product at these coefficients is
\begin{align*}
 \left(\sum_{r=1}^M g(t_r)\right)
 \left(\sum_{r=1}^M t_rb_r^2\right)
 &=
 (4M-3k)\left(Q_k+\frac{Q-Q_k}{3}\right)\\
 &=
 \left(\frac43-\theta\right)(MQ+2MQ_k).
\end{align*}

Recall that the construction uses $\lfloor M/2\rfloor$ bags of volume $3/(4M)$ and the same number of bags of volume~$5/(4M)$, together with one bag of volume $1/M$ if $M$ is odd. Direct summation gives
\[
 MQ\le\frac{17}{16},
 \qquad
 MQ_k\le
 \begin{cases}
  9\theta/16, & \theta\le1/2,\\
  25\theta/16-1/2, & \theta\ge1/2.
 \end{cases}
\]
For even $M$, both bounds hold with equality. For odd $M$, we have $MQ=17/16-1/(16M)$. The bound for~$MQ_k$ is an equality when $\theta\le1/2$, while for $\theta\ge1/2$ its value is $1/(16M)$ below the displayed upper bound. Thus, the same bounds apply in both cases.

It remains to bound the two resulting quadratics.
Completing the square gives
\begin{align*}
 \left(\frac43-\theta\right)
 \left(\frac{17}{16}+\frac{9\theta}{8}\right)
 &=
 \frac{1681}{1152}
 -\frac98\left(\theta-\frac7{36}\right)^2,\\
 \left(\frac43-\theta\right)
 \left(\frac1{16}+\frac{25\theta}{8}\right)
 &=
 \frac{41209}{28800}
 -\frac{25}{8}
  \left(\theta-\frac{197}{300}\right)^2.
\end{align*}
Since $41209/28800<1681/1152$, both expressions are at most $1681/1152$. Applying~\eqref{eq:imbalanced-product} therefore yields
\[
 R(s)\le\frac{1681}{1152}.
\]
\qed
\end{proof}

\section{Additional variants (mentioned in \Cref{sec:results})}\label{app:variants}

\subsection{The non-adaptive model}
\label{app:non-adap-detail}
We consider the non-adaptive variant of our two-stage scheduling problem for an unknown number of identical machines. An instance consists of $n$ jobs with processing times $p_j>0$, $j\in[n]$, and the number $M$ of bags. All jobs are available at time zero.

In Phase~I, the algorithm partitions the jobs into $M$ bags, some of which may be empty. An adversary then reveals the number $m\in[M]$ of available unit-speed machines. In Phase~II, the algorithm assigns every bag as a whole to a machine and chooses the order of the bags on each machine. Within each bag, jobs are processed in SPT order, which is optimal for any assignment and ordering of the bags. The difference from the original model is that bags must be scheduled contiguously.

For a fixed instance, let $\alg(m,M)$ denote the total completion time of the algorithm's schedule and let~$\opt_m$ denote the minimum total completion time when the number $m$ of machines is known in advance and jobs may be assigned individually, without bag constraints. An algorithm is $\rho$-robust if $\alg(m,M)\le\rho\,\opt_m$ for every instance and every $m\in[M]$. The optimal robustness factor is the infimum of the robustness factors of all deterministic algorithms. 

Let $k$ denote the number of distinct processing times. Writing $M+k-1=kr+s$, where $r\ge1$ and~$0\le s<k$, we~define
\[
 \rho_{M,k}:=
 \frac{M}{k}\left(\frac{s}{r+1}+\frac{k-s}{r}\right).
\]
For $M,k\ge2$, we show that $\rho_{M,k}$ is the optimal deterministic robustness factor for~$M$ bags and instances with $k$ distinct processing times. The lower bound is proved in~\Cref{thm:non-adap-LB}, and a matching algorithm is given in~\Cref{thm:nonadaptive-upper}.

Since $s(r+1)+(k-s)r=M+k-1$ and $r=\lceil M/k\rceil$, the Cauchy--Schwarz inequality and the definition of $\rho_{M,k}$ give
\[
 \frac{Mk}{M+k-1}
 \le \rho_{M,k}
 \le \frac{M}{r}
 \le \min\{M,k\}.
\]

For the lower bound, the adversary only needs to choose between one and $M$ available machines. We construct job classes with increasingly separated processing times. On one machine, processing a fixed positive fraction of a longer class before a fixed positive fraction of a shorter class leads to a total completion time that is unbounded compared to that of SPT. Since bags must be processed contiguously, bounded robustness on one machine forces the limiting assignment to respect SPT order across the bags. This restricts how widely classes can be distributed among bags.
When all $M$ machines are available, however, an unrestricted schedule can distribute every class evenly. Comparing the resulting costs yields the lower bound.

Towards this, we first establish a structural property. Index the processing-time classes $J_1,\ldots,J_k$ in increasing order, and let $x_{ib}$ denote the fraction of class $J_i$ assigned to bag $B_b$, with $\sum_{b=1}^M x_{ib}=1$ for every~$i\in[k]$. The following lemma bounds the sum, over all classes, of the number of bags containing a positive fraction of that class. In the lower-bound proof, we apply the lemma to the limiting assignment.

\begin{lemma}\label{lemma:Number_Packages_Ordered_Bags}
Assume that the bags are ordered such that $x_{ja}x_{ib}=0$ for every $i<j$ and~$a<b$.
Denote by
\begin{align*}
z_i:=\bigl|\{b\in[M]:x_{ib}>0\}\bigr|
\end{align*}
the number of bags containing a positive fraction of class $J_i$. Then
\begin{align*}
    \sum_{i=1}^k z_i\leq M+k-1.
\end{align*}
\end{lemma}
\begin{proof}
For every class $J_i$, let
\begin{align*}
    a_i:=\min\{b\in[M]:x_{ib}>0\},
    \qquad
    b_i:=\max\{b\in[M]:x_{ib}>0\}
\end{align*}
denote the first and last bags containing jobs from $J_i$, respectively.
The assumption implies $b_i\leq a_{i+1}$ for every~$i\in[k-1]$, since otherwise $x_{i+1,a_{i+1}}x_{i,b_i}>0$.

Since only bags between $a_i$ and $b_i$ can contain jobs from $J_i$, we get $z_i\leq b_i-a_i+1$. Summing over all classes~yields
\begin{align*}
    \sum_{i=1}^k z_i
    &\leq \sum_{i=1}^k(b_i-a_i+1)
    =k+b_k-a_1+\sum_{i=1}^{k-1}(b_i-a_{i+1})
    \leq M+k-1,
\end{align*}
where the last inequality follows from $a_1\geq1$, $b_k\leq M$, and $b_i\leq a_{i+1}$ for every $i\in[k-1]$.\qed
\end{proof}

We now construct instances for which bounded robustness on one machine forces the ordering condition of~\Cref{lemma:Number_Packages_Ordered_Bags} in the limit.

\begin{theorem}\label{thm:non-adap-LB}
Let $M,k\ge1$. For every deterministic non-adaptive algorithm using $M$ bags and every $\varepsilon>0$, there exists an instance with exactly $k$ distinct processing times on which the adversary can force a ratio of at least $\rho_{M,k}-\varepsilon$ by choosing either one or~$M$ identical machines.
\end{theorem}
\begin{proof}
The claim is immediate if $M=1$ or $k=1$, since $\rho_{M,k}=1$. Therefore, assume~$M,k\ge2$, and fix a deterministic algorithm and $\varepsilon>0$.

For every integer $T\ge2$, let $q=MT$ and construct an instance with $k_i=T^{k-i}q$ jobs of processing time~$\ell_i=T^{2(i-1)}$ for every $i\in[k]$. Then $\ell_1<\cdots<\ell_k$, and $M\mid k_i$ for every $i$. Write $\alg_T(m,M)$ and $\opt_{m,T}$ for the algorithm's cost and the unrestricted optimum on this instance, respectively. All asymptotic statements below refer to $T\to\infty$, with $M$ and $k$ fixed.

Let $W:=T^{2k-2}q^2$. By construction, we have $\ell_i k_i^2=W$ for every class $i\in[k]$.
Moreover, the linear terms satisfy
\begin{align*}
    \ell_i k_i &= \frac{W}{k_i} \leq \frac{W}{q}=o(W).
\end{align*}
For every $i<j$, the cross-contribution in SPT order satisfies
\begin{align}\label{eq:Equal_Area_Lower_Order}
    \ell_i k_i k_j &= WT^{i-j} \leq \frac{W}{T} =o(W).
\end{align}
Thus, all classes have the same leading contribution, while the linear terms and cross-contributions in SPT order are of lower order.

Each pair in which a job of class~$J_j$ precedes a job of class~$J_i$, where $i<j$, incurs an additional cost of~$\ell_j-\ell_i$ compared with SPT order. Moreover,
\begin{equation}\label{eq:Additional_Costs_Inversion_Infty}
 \frac{(\ell_j-\ell_i)k_i k_j}{W}
 =T^{j-i}-T^{i-j}
 \longrightarrow\infty.
\end{equation}
Hence, a fraction bounded away from zero of such inverted pairs incurs an unbounded cost relative to $W$.

Suppose, for contradiction, that the claimed lower bound does not hold for this algorithm and $\varepsilon$. Then, for every integer $T\ge2$ and every $m\in\{1,M\}$,
\begin{equation}\label{eq:LB_Contradiction_Assumption}
 \frac{\alg_T(m,M)}{\opt_{m,T}}
 <\rho_{M,k}-\varepsilon.
\end{equation}
For every $T$, relabel the bags $B_1,\ldots,B_M$ according to the order in which the algorithm processes them when~$m=1$. Let $x_{ib}^{(T)}$ denote the fraction of class~$J_i$ assigned to bag~$B_b$. Thus,
\[
 x_{ib}^{(T)}\ge0,
 \qquad
 \sum_{b=1}^M x_{ib}^{(T)}=1.
\]

Since the vectors $(x_{ib}^{(T)})_{i\in[k],b\in[M]}$ lie in the compact set $[0,1]^{kM}$, we may pass to a subsequence of integers~$T\to\infty$ such that
\begin{equation}\label{eq:LB_Limit_Vector}
 x_{ib}^{(T)}\longrightarrow x_{ib}^*
\end{equation}
for every $i\in[k]$ and $b\in[M]$. The limiting fractions satisfy $x_{ib}^*\ge0$ and $\sum_b x_{ib}^*=1$. In the remainder, all limits are taken along this subsequence.

First, let $m=1$. Since an optimal schedule processes the jobs in SPT order, the preceding estimates yield
\[
 \opt_{1,T}=\frac{kW}{2}+o(W).
\]
Suppose that $x_{ja}^*x_{ib}^*>0$ for some $i<j$ and $a<b$. Every job of class~$J_j$ in bag~$B_a$ precedes every job of class~$J_i$ in bag~$B_b$. By
\eqref{eq:Additional_Costs_Inversion_Infty},
\[
 \frac{\alg_T(1,M)-\opt_{1,T}}{W}
 \ge
 \bigl(T^{j-i}-T^{i-j}\bigr)
 x_{ja}^{(T)}x_{ib}^{(T)}
 \longrightarrow\infty.
\]
This contradicts \eqref{eq:LB_Contradiction_Assumption}, since $\opt_{1,T}/W\to k/2$. Thus, $x_{ja}^*x_{ib}^*=0$ for every $i<j$ and~$a<b$.

Let
\[
 z_i:=\bigl|\{b\in[M]:x_{ib}^*>0\}\bigr|
\]
denote the number of bags containing a positive fraction of class~$J_i$ in the limiting assignment. By~\Cref{lemma:Number_Packages_Ordered_Bags},
\begin{equation}\label{eq:LB_Total_Support}
 \sum_{i=1}^k z_i\le M+k-1.
\end{equation}

Now, let $m=M$. Since $M\mid k_i$ for every $i$, we may assign exactly $k_i/M$ jobs of each class to every machine and process them in SPT order. Hence,
\begin{align}
    \opt_{M,T}
    &\leq
    \sum_{i=1}^k
    \left(
        \frac{\ell_i k_i^2}{2M}
        +\frac{\ell_i k_i}{2}
    \right)
    +\frac1M\sum_{1\le i<j\le k}\ell_i k_i k_j
    \notag\\
    &=\frac{kW}{2M}+o(W).
    \label{eq:OPT_LB_Beweis}
\end{align}

The quadratic self-contribution of class~$J_i$ within bag~$B_b$ is
\[
 \frac{\ell_i}{2}
 \bigl(k_i x_{ib}^{(T)}\bigr)^2
 =\frac W2\bigl(x_{ib}^{(T)}\bigr)^2.
\]
All remaining contributions are nonnegative, independently of how the bags are assigned and ordered in Phase~II. Therefore,
\[
 \alg_T(M,M)
 \ge
 \frac W2\sum_{i=1}^k\sum_{b=1}^M
 \bigl(x_{ib}^{(T)}\bigr)^2.
\]
Together with \eqref{eq:LB_Limit_Vector} and \eqref{eq:OPT_LB_Beweis}, this yields
\[
 \liminf_{T\to\infty}
 \frac{\alg_T(M,M)}{\opt_{M,T}}
 \ge
 \frac Mk\sum_{i=1}^k\sum_{b=1}^M(x_{ib}^*)^2.
\]
For every $i\in[k]$, the Cauchy--Schwarz inequality gives
\[
 1
 =\left(\sum_{b:x_{ib}^*>0}x_{ib}^*\right)^2
 \le z_i\sum_{b=1}^M(x_{ib}^*)^2.
\]
Consequently,
\[
 \liminf_{T\to\infty}
 \frac{\alg_T(M,M)}{\opt_{M,T}}
 \ge \frac Mk\sum_{i=1}^k\frac1{z_i}.
\]

It remains to minimize $\sum_i 1/z_i$ over integers $1\le z_i\le M$ satisfying \eqref{eq:LB_Total_Support}. Since $1/z$ is decreasing and~$M+k-1\le kM$, the minimum is attained when $\sum_i z_i=M+k-1$. By convexity, in a minimizing choice the values $z_i$ differ by at most one. Thus, such a choice has exactly $s$ values equal to $r+1$ and $k-s$ values equal to $r$, where $M+k-1=kr+s$ and~$0\le s<k$. It follows that
\[
 \sum_{i=1}^k\frac1{z_i}
 \ge \frac{s}{r+1}+\frac{k-s}{r}.
\]
Hence,
\[
 \liminf_{T\to\infty}
 \frac{\alg_T(M,M)}{\opt_{M,T}}
 \ge
 \frac Mk\left(\frac{s}{r+1}+\frac{k-s}{r}\right)
 =\rho_{M,k}.
\]
This contradicts \eqref{eq:LB_Contradiction_Assumption} and proves the claim. \qed
\end{proof}

For $k\geq2$, the lower bound can be matched by the following algorithm. In Phase~I, sort all jobs in SPT order and partition this sequence into $M$ consecutive bags, allowing empty bags, to minimize their total completion time when processed on separate unit-speed machines. In Phase~II, choose a bag assignment whose cost is at most the uniform average over all assignments in which the numbers of bags per machine differ by at most one. Here, empty bags are counted, and the bags on each machine are processed in their order along the SPT sequence. We call this algorithm~\textsc{SPT-Cut}.

Both phases can be implemented in polynomial time. For Phase~I, a dynamic-programming state records a prefix of the SPT sequence and the number of bags. A transition chooses the beginning of the last bag, which may be empty. After precomputing the cost of every consecutive block of jobs, this gives a running time of $O(Mn^2)$.

For Phase~II, fix balanced bag counts $h_1,\ldots,h_m$ and consider a uniformly random assignment respecting these counts. By symmetry of the identical machines, its expected cost equals the average over all balanced assignments. We assign the bags successively, each time choosing a machine with a free bag position that minimizes the conditional expected final cost. Since the current conditional expectation is a weighted average over the possible choices, this step never increases it. The final assignment therefore has cost at most the initial average.

To compute these conditional expectations, observe that the contribution within each bag is fixed. For two bags~$B_a,B_b$ with $a<b$, assigning them to the same machine adds the contribution
\[
 w_{ab}:=\left(\sum_{j\in B_a}p_j\right)|B_b|.
\]
Suppose that $R$ bags remain unassigned and machine~$i$ has $d_i$ free bag positions. An unassigned bag shares a machine with a bag already assigned to machine~$i$ with probability~$d_i/R$. Two unassigned bags share a machine with probability
\[
 \frac{\sum_{i=1}^m d_i(d_i-1)}{R(R-1)}
 \qquad\text{if }R\ge2.
\]
Thus, each conditional expected cost can be computed by summing the fixed contributions within bags and the expected contributions between pairs of bags.

On one machine, processing the bags in their original order preserves the unrestricted SPT order and is therefore optimal. When all $M$ machines are available, the chosen cuts minimize the cost among partitions into consecutive bags. For intermediate numbers of machines, averaging over balanced assignments bounds the additional delay caused by 
processing several bags on the same machine.

\begin{theorem}\label{thm:nonadaptive-upper}
For $M,k\ge2$, algorithm \textsc{SPT-Cut} is $\rho_{M,k}$-robust in the non-adaptive model with $M$ bags and an unknown number of identical machines, on instances with $k$ distinct processing times.
\end{theorem}

\begin{proof}
Fix an instance and a revealed number $m\in[M]$ of machines. Let $J_i$ contain $k_i$ jobs of processing time $\ell_i$, where
$\ell_1<\cdots<\ell_k$, and put
\[
 c:=\frac{\rho_{M,k}}M.
\]
By the bounds on $\rho_{M,k}$, we have $0<c\le1$.

We distinguish the jobs' own processing times from the delays caused by other jobs. More precisely, let
\[
 P:=\sum_{i=1}^k\ell_i k_i,
 \qquad
 K:=\sum_{i=1}^k\ell_i\binom{k_i}{2}
      +\sum_{i<j}\ell_i k_i k_j.
\]
Thus, the total completion time of the one-machine SPT schedule is $P+K$. Each pair of jobs contributes the processing time of its shorter job to $K$.

Let $C$ denote the total completion time when each bag constructed in Phase~I is processed on a separate unit-speed machine. We first show that
\begin{equation}\label{eq:nonadaptive-cut-cost}
 C\le P+cK.
\end{equation}
For this, we construct a random comparison partition of the SPT sequence into $M$ consecutive bags. Splitting class~$J_i$ into $z_i$ parts uses $z_i-1$ cuts. Thus, the $M-1$ available cuts give the same sum $\sum_i z_i=M+k-1$, the same total that appears in the lower-bound minimization. We distribute these parts as evenly as possible among the classes.

Recall that $M+k-1=kr+s$, where $0\le s<k$. Choose $s$ classes uniformly at random and set $z_i=r+1$ for these classes and $z_i=r$ for the others. Then
\[
 \sum_{i=1}^k z_i=M+k-1,
 \qquad
 \mathbb{E}\left[\frac1{z_i}\right]
 =\frac1k\left(\frac{s}{r+1}+\frac{k-s}{r}\right)
 =c
\]
for every $i\in[k]$.

Split each class $J_i$ into $z_i$ parts whose sizes differ by at most one, and order these parts uniformly at random within the class. Place a cut between consecutive parts of the same class. This gives $\sum_i(z_i-1)=M-1$ cuts and hence $M$ consecutive bags. Observe that parts and bags may be empty. In particular, coinciding cuts create empty bags.

Consider first the contribution of pairs within a class $J_i$. If its part sizes are $a_1,\ldots,a_{z_i}$, then
\[
 \sum_{h=1}^{z_i}\binom{a_h}{2}
 \le
 \frac{k_i}{2}
 \left(\left\lceil\frac{k_i}{z_i}\right\rceil-1\right)
 \le
 \frac1{z_i}\binom{k_i}{2}.
\]
Thus, the expected waiting-time contribution within this class is at most $c\ell_i\binom{k_i}{2}$.

Now consider two classes $J_i$ and $J_j$ with $i<j$. Only the last part of $J_i$ can share a bag with jobs of $J_j$. Conditional on $z_i$, this part has expected size $k_i/z_i$. Hence, the expected contribution between these classes is at most
\[
 \mathbb{E}\left[\frac{\ell_i k_i k_j}{z_i}\right]
 =c\ell_i k_i k_j.
\]
The contribution $P$ of the jobs' own processing times remains unchanged. Summing over all classes and pairs of classes, the expected cost of the comparison partition is at most $P+cK$. Since Phase~I chooses a minimum-cost partition, \eqref{eq:nonadaptive-cut-cost} follows.

We next consider Phase~II. Since the bags are consecutive parts of the SPT sequence, processing them in their original order gives an SPT schedule on each machine. Contributions within bags are always paid, while a contribution between two different bags is paid precisely when these bags share a machine. We aim to bound the probability that two fixed bags share a machine under a uniformly random balanced assignment.

Let $\beta$ be the probability that two fixed distinct bags are assigned to the same machine. If the machine bag counts are $h_1,\ldots,h_m$, then
\[
 \beta
 =\frac{\sum_{a=1}^m h_a(h_a-1)}{M(M-1)}
 \le
 \frac{\lceil M/m\rceil(M-m)}{M(M-1)}
 \le
 \frac{2(M-m)}{m(M-1)}.
\]
Here, we used $1\le h_a\le\lceil M/m\rceil$, $\sum_a(h_a-1)=M-m$, and $\lceil M/m\rceil\le2M/m$.

The algorithm chooses an assignment whose cost is at most this average. Therefore,
\begin{align}
 \alg(m,M)
 &\le C+\beta(P+K-C)
 \notag\\
 &\le P+\bigl(c+(1-c)\beta\bigr)K,
 \label{eq:nonadaptive-assignment-cost}
\end{align}
where the second inequality follows from
\eqref{eq:nonadaptive-cut-cost} and $\beta\le1$.

It remains to compare this bound with the optimum. Let $\ell_0=0$ and $N_i:=\sum_{j=i}^k k_j$. Using the layer decomposition from the proof of~\Cref{lem:prefix-transfer}, we obtain
\[
 P=\sum_{i=1}^k(\ell_i-\ell_{i-1})N_i,
 \qquad
 K=\sum_{i=1}^k(\ell_i-\ell_{i-1})\binom{N_i}{2}.
\]
In each layer, the machine job counts sum to $N_i$. By Cauchy--Schwarz, their squared counts sum to at least~$N_i^2/m$. Consequently,
\begin{align}
 \opt_m
 &\ge
 \sum_{i=1}^k(\ell_i-\ell_{i-1})
 \left(\frac{N_i^2}{2m}+\frac{N_i}{2}\right)
 \notag\\
 &=\frac Km+\frac{m+1}{2m}P.
 \label{eq:nonadaptive-opt-lower}
\end{align}
Finally, Cauchy--Schwarz and $k\ge2$ imply
\[
 c\ge\frac{k}{M+k-1}\ge\frac2{M+1}.
\]
In particular, $2(1-c)/(M-1)\le c$, and hence
\begin{align*}
 c+(1-c)\beta
 &\le c+\frac{2(1-c)(M-m)}{m(M-1)}\\
 &\le c+\frac{c(M-m)}m
 =\frac{\rho_{M,k}}m.
\end{align*}
Moreover, since $m\le M$,
\[
 \rho_{M,k}=Mc
 \ge\frac{2M}{M+1}
 \ge\frac{2m}{m+1}.
\]
Combining these inequalities with \eqref{eq:nonadaptive-assignment-cost} and \eqref{eq:nonadaptive-opt-lower}, we obtain
\begin{align*}
 \alg(m,M)
 &\le P+\frac{\rho_{M,k}}mK\\
 &\le
 \rho_{M,k}
 \left(\frac Km+\frac{m+1}{2m}P\right)\\
 &\le \rho_{M,k}\opt_m.
\end{align*}
This holds for every $m\in[M]$ and proves the claim. \qed
\end{proof}

For $k\ge M$, we have $r=1$ and $s=M-1$, so
\[
 \rho_{M,k}=M-\frac{M(M-1)}{2k}.
\]
Thus, for fixed $M$, the optimal robustness factor converges to $M$ as~$k$ increases. Conversely, for fixed $k \geq 2$, it approaches~$k$ as~$M$ increases. Without restricting the number of distinct processing times, the optimal robustness factor for $M$ bags is therefore $M$. This is attained even by the trivial algorithm that places all jobs in SPT order into a single bag.

For $k=1$, the formula gives only the trivial lower bound $\rho_{M,1}=1$. All jobs have the same processing time, so requiring bags to be processed contiguously does not change the cost of any bag assignment. Thus, this case coincides with the bricks setting considered in \Cref{sec:Equal_Speeds}.

\subsection{A stronger adversary: failures after job assignment}
\label{app:remarks-proofs}
Liu, Zhang, and Zhao~\cite{DBLP:conf/ijcai/LiuZZ26} study machine failures for makespan minimization under a weak and a strong adversary.
In their model, jobs are assigned directly to machines in Phase~I, and up to $k$ machines may fail afterwards.
A strong adversary determines which machines fail, whereas a weak adversary determines only their number, and the algorithm chooses which machines fail.
Jobs on surviving machines cannot be reassigned, while jobs from failed machines may be reassigned individually.
We adopt this model for total completion time, where the jobs on each machine may be reordered in Phase~II, and allow up to $M-1$ failures. For general speeds, the adversary chooses the speed of each machine after Phase~I, and only jobs on machines of speed zero may be reassigned.

We first show that extending the strong-adversary model to speed-robust scheduling with general machine speeds leads to unbounded robustness, already for $M=2$.
\begin{theorem}
    For general speeds, no deterministic algorithm for minimizing the total completion time of sand under a strong adversary achieves a bounded robustness factor.
\end{theorem}
\begin{proof}
    Choose $M=2$ and let $y_1\geq y_2\geq0$ with $y_1+y_2=1$ denote the machine loads after Phase~I. The adversary chooses speeds $s_1=\varepsilon$ and $s_2=1-\varepsilon$ for some $\varepsilon\in(0,1)$.

    Since both speeds are positive, no machine fails and no reassignment is allowed.
    By~\Cref{sec:prelim}, we have~$\alg(s)=(y_1^2/s_1+y_2^2/s_2)/2$ and, since $s_1+s_2=1$, $\opt(s)=1/2$. Moreover, $y_1\geq1/2$, and thus the robustness ratio satisfies
    \begin{align*}
        \frac{\alg(s)}{\opt(s)}
        &=\frac{y_1^2}{s_1}+\frac{y_2^2}{s_2}
        \geq\frac{y_1^2}{\varepsilon}
        \geq\frac{1}{4\varepsilon}.
    \end{align*}
    As $\varepsilon\to0$, the ratio becomes arbitrarily large, proving the claim.\qed
\end{proof}
We therefore restrict attention to machine speeds in $\{0, 1\}$, for which an optimal algorithm exists.
\begin{theorem}\label{theorem:Opt_For_Strong_ADV}
    For speeds in $\{0,1\}$, there exists an optimal deterministic algorithm for minimizing the total completion time of bricks, rocks, and sand under a strong adversary.
\end{theorem}
\begin{proof}
    For sand, distribute the volume equally among the $M$ machines in Phase I. If $m$ machines survive, distribute the volume of failed machines equally among them. Every surviving machine then has load $1/m$, which is optimal.

    For bricks and rocks, index the jobs such that $p_1\leq\ldots\leq p_n$.
    In Phase~I, assign the jobs $1,\ldots,n$ cyclically to the $M$ machines.
    In Phase~II, fix an arbitrary set of surviving machines, and let $m\leq M$ denote its size.
    For $i=1,\ldots,\lceil n/m\rceil$, let $G_i$ be the set of jobs~$j$ with $\lceil(n-j+1)/m\rceil=i$, i.e., the $i$-th group of $m$ jobs in non-increasing order of size (the last group may be smaller).
    
    Since two jobs assigned to the same machine in Phase I have indices differing by at least $M$, and each group consists of at most $m\leq M$ consecutive jobs, no machine contains two jobs from the same group.
    For each group $G_i$, consider the jobs of the failed machines. Assign each of these jobs to an available machine that does not yet contain a job from $G_i$.
    This is always possible since $|G_i|\leq m$ and the jobs already assigned to available machines occupy distinct machines. Thus, each full group has exactly one job on each available machine, while the last group has at most one job per machine. On each machine, process the jobs in SPT order and, among jobs of equal size, the job from the group with larger index first.
    
    It remains to show that the resulting schedule is optimal.
    An optimal schedule on $m$ machines processes each machine in SPT order. In such a schedule, a job~$j$ contributes~$a_jp_j$ to the objective, where $a_j$ is the number of jobs processed no earlier than job~$j$ on its machine, including~$j$ itself.

    Since each coefficient occurs at most once per machine, it occurs at most $m$ times in total. By the rearrangement inequality, the objective is therefore bounded from below by assigning coefficient $1$ to the $m$ largest jobs, coefficient $2$ to the next $m$ largest jobs, and so on.
    This yields
    \begin{align*}
        \opt_m \geq
        \sum_{j=1}^n
        \left\lceil\frac{n-j+1}{m}\right\rceil p_j.
    \end{align*}
    In our schedule, each job in $G_i$ is followed by exactly $i-1$ jobs on its machine, one from each preceding group. Hence, its coefficient is $i$.
    Consequently,
    \begin{align*}
        \alg(m,M)
        &=\sum_{i=1}^{\lceil n/m\rceil} i\sum_{j\in G_i}p_j\\
        &=\sum_{j=1}^n \left\lceil\frac{n-j+1}{m}\right\rceil p_j
        \leq\opt_m.
    \end{align*}
    Since $\alg(m,M)\geq\opt_m$ holds trivially, the algorithm is optimal.\qed
\end{proof}

Finally, we consider an \textit{indivisible strong adversary}.
Here, the adversary decides which machines fail, and all jobs from the same failed machine must be reassigned together to a single surviving machine.
For general machine speeds, the unboundedness result carries over to the indivisible strong adversary, since the adversary may also choose not to fail any machine.
We now establish an upper bound for sand using the algorithm that distributes the volume equally among all machines.
Afterward, we derive lower bounds for bricks, rocks, and sand.
\begin{theorem}
    For speeds in $\{0,1\}$, the algorithm that distributes sand equally among the machines in Phase~I and optimally reassigns the volumes of failed machines in Phase II is $9/8$-robust for minimizing the total completion time under an indivisible strong adversary.
    This bound is tight for the algorithm.
\end{theorem}
\begin{proof}
    Let $M$ be the number of initial machines and let $m\leq M$ machines remain available.
    In Phase~I, each machine receives volume $1/M$. Since the volumes of failed machines are reassigned as a whole, every surviving machine ends up with load $k_i/M$, where $k_i\geq1$ is the number of bags it holds and $\sum_{i=1}^m k_i=M$.
    As $\sum_i k_i^2$ is minimized by balanced vectors, an optimal reassignment yields $k_i\in\{\lfloor M/m\rfloor,\lceil M/m\rceil\}$.
    
    Write $M=qm+r$ with $0\leq r<m$. Thus, $m-r$ machines have final load $q/M$, while $r$ machines have load~$(q+1)/M$.
    Since $\opt_m=1/(2m)$, the robustness ratio is
    \begin{align*}
        R(m)
        &=\frac{m}{M^2}\left((m-r)q^2+r(q+1)^2\right)
        =1+\frac{r(m-r)}{M^2}.
    \end{align*}
    Since $q\geq1$, we have $M\geq m+r$. Moreover, $(m+r)^2-8r(m-r)=(m-3r)^2\geq0$, and hence
    \begin{align*}
        R(m)
        \leq1+\frac{r(m-r)}{(m+r)^2}
        \leq1+\frac18
        =\frac98.
    \end{align*}
    The bound is attained for $M=4$ and $m=3$, where the final machine loads are $(2,1,1)/4$ and $R(m)=9/8$.\qed
\end{proof}

\begin{theorem}\label{theorem:LB_Indisible_Strong_ADV}
    For speeds in $\{0,1\}$, no deterministic algorithm for minimizing the total completion time under an indivisible strong adversary achieves a robustness factor smaller than $7/6$ for bricks or rocks, or~$(60-6\sqrt{2})/49>1.0513$ for sand.
\end{theorem}
\begin{proof}
    For bricks, consider an instance with four unit-size jobs and three machines. If at least one machine receives no job, the adversary keeps all three machines available. The algorithm incurs a cost of at least $6$, whereas the optimum has cost $5$, yielding a ratio of at least $6/5>7/6$.

    Otherwise, two machines receive one job, while one machine receives two jobs. The adversary lets the machine with two jobs fail. Hence, the machine loads are $(3,1)$ with cost $7$, while the optimum $(2,2)$ has cost $6$. Thus, the ratio is $7/6$. Since bricks are a special case of rocks, the same lower bound applies to rocks.

    For sand, consider an instance with three machines. Let $a\geq b\geq c\geq 0$ denote their volumes on the three machines with $a+b+c=1$.
    The adversary considers two scenarios. First, if all machines remain available, no reassignment is allowed. Using~$a+c=1-b$, we obtain a lower bound on the ratio solely in terms of the median load $b$:
    \begin{align*}
        R(3)=3(a^2+b^2+c^2)&=1+18\left(b-\frac13\right)^2+6(a-b)(b-c)\\
        &\ge 1+18\left(b-\frac13\right)^2.
    \end{align*}
    Second, the adversary lets the machine with load $a$ fail. Since its load must be reassigned as a whole, assigning it to the less loaded surviving machine is optimal. Thus,
    \begin{align*}
        R(2)\geq 2(b^2+(a+c)^2)=1+(1-2b)^2.
    \end{align*}
    Let $\beta=\frac{4+\sqrt{2}}{14}$ and $\rho^\star=\frac{60-6\sqrt{2}}{49}$.
    If $b\le\beta$, then $R(2)\ge 1+(1-2\beta)^2=\rho^\star$.
    Otherwise, we have $b>\beta>1/3$, implying $R(3)\ge 1+18\left(\beta-\frac13\right)^2=\rho^\star$.
    
    Thus, the adversary can always enforce a robustness ratio of at least $\rho^\star$. For $M=3$, this bound is best possible: consider the initial distribution $(a,b,c)=(\beta,\beta,1-2\beta)$. If all machines remain available, then $R(3)=\rho^\star$. If exactly one machine fails, an optimal reassignment yields loads $(\beta,1-\beta)$, giving $R(2)=1+(1-2\beta)^2=\rho^\star$. If only one machine survives, the ratio is one. Hence, this distribution achieves worst-case robustness $\rho^\star$.
    \qed
\end{proof}

\end{document}